\documentstyle[aps]{revtex}

\def \bfgr #1{ \mbox {{\boldmath $#1$}}}
\def \dfrac #1#2 {\displaystyle\frac{#1}{#2}}
\def\Upp{u^{+}(p_0,|{\bf p}|)}
\def\Wpp{w^{+}(p_0,|{\bf p}|)}

\newcommand{\bqn}{\begin{small}\begin{eqnarray}}
\newcommand{\eqn}{\end{eqnarray}\end{small}}
\newcommand{\beq}{\begin{equation}}
\newcommand{\eeq}{\end{equation}}

\newcommand{\mbf}[1]{\mbox{\boldmath$#1$}}
\newcommand{\nrmp}{|\mbox{\boldmath $p$}|}

\def\Pe{^1P_1^{e}}
\def\Po{^1P_1^{o}}

\begin{document}
 
\title{ 
Bethe-Salpeter Amplitudes and \\[3mm]
Static Properties of the Deuteron  }

\author{ 
L.P.  Kaptari$^{a,b}$,
A.Yu.  Umnikov$^c$,
S.G.  Bondarenko$^b$,
K.Yu.  Kazakov$^d$, \\
F.C.  Khanna$^{e,f}$,
and
B. K\"ampfer$^{a,g}$}

\address{
$^a$Research Center Rossendorf,
Institute for Nuclear and Hadron Physics, \\
PF 510119 01314, Dresden,Germany \\[1mm]
$^b$Bogoliubov Laboratory of Theoretical Physics, JINR, \\
141980 Dubna, Moscow Region, Russia \\[1mm]
$^c$INFN Section Perugia, via A. Pascoli, I-06100,Italy \\[1mm]
$^d$Far Eastern State University, Vladivostok, 690000, Russia\\[1mm]
$^e$University of Alberta, Edmonton, Alberta T6G 2J1, Canada,\\[1mm]
$^f$TRIUMF, 4004 Wesbrook Mall, Vancouver, BC, Canada, V6T 2A3\\[1mm]
$^g$Institute for Theoretical Physics, Technical University,
01062 Dresden, Germany}
 
\maketitle

\begin{abstract}
Extended calculations of the deuteron's static properties,
based on the numerical solution of the Bethe-Salpeter
equation, are presented.
A formalism is developed, which provides a comparative analysis
of the covariant amplitudes in various representations
and nonrelativistic wave functions.
The magnetic and quadrupole moments of
the deuteron are calculated in the Bethe-Salpeter formalism
and the r\^ole of relativistic corrections is discussed.
\end{abstract}

\section{Introduction}
A theory applicable for studying nuclear phenomena, involving momentum
transfers of a few GeV or higher, must be relativistic.  A traditional
approach to processes with nuclei, based on the nonrelativistic
Schr\"odinger wave functions, is not adequate if a large momentum
transfer "sneaks" into the nuclear amplitudes, and the corresponding
nucleon momentum $p$ becomes large, say $p \geq m$ ($m$ is the nucleon mass).
One can extend the usage of  nonrelativistic wave functions by
incorporating successively the relativistic corrections $\sim (p/m)^n$,
however, it will eventually fail at some point of $p$.
On the other hand, the
nonrelativistic approach was the only one which allowed for the
detailed description of the static properties of the nuclei and
low and intermediate energy nuclear reactions.

In recent decades, extensive studies of few-nucleon
systems were performed within Lorentz invariant
models \cite{tjon,gross,gross2,coester,karman}. The success of these
elaborate studies allows one to conclude that the covariant approach
has now the capability to replace, at least for
few-nucleon systems, the approaches relying on nonrelativistic wave
functions \cite{gross-ams}.  Most of the phenomenological success in
the relativistic treatment of few-nucleon systems has been
achieved within such models which are
based on the covariant meson-nucleon theory
and dynamical equations \cite{tjon,gross,gross2}.  In these models,
the satisfactory results have been obtained for
the nucleon-nucleon scattering,
the properties of the lightest nuclei, various electromagnetic and hadronic
interactions with nuclei, and some advance have been achieved for
many-body nuclear systems (see e.g. discussions and references in
ref.\ \cite{gross2}).

The deuteron, as the simplest nuclear system, is an appealing object
to be described by the models invented in the realm of nuclear physics.
At the same time, there is a fair amount of  experimental information
available about the deuteron's properties themselves
and reactions with the deuteron.
More interesting and precise data is expected after the start of
the exciting research program at CEBAF. Therefore, there is a
possibility to compare  exact theoretical results with the
experimental data in a clear way, not dimmed by extra
effects, such as the ``more-then-two-body" calculations.

Still, the relativistic approach to the deuteron is not as popular as
the one utilizing nonrelativistic wave functions \cite{paris,bonn}.
There are seemingly two main reasons for this.
{\em First},
the deuteron, as any other nucleus, is
essentially a nonrelativistic system, since it is composed of weakly
bound massive nucleons. The bulk of the static properties of such a
system obviously can be fitted in the nonrelativistic approach by
adjusting the phenomenological potential or the wave function.
Besides, the experimental data for the reactions with the deuteron is
also mainly available in the nonrelativistic domain.
{\em Second},
the relativistic models, especially those based on  field theory,
are technically more difficult and have a more sophisticated physical
interpretation than the nonrelativistic approach.  Both these reasons,
together, define the typical pattern for the attempts devoted to promote
the relativistic description of the nuclei.  These works are usually
highly specified for the particular reactions or kinematic domains where the
advantage of the covariant approach can be explicitly displayed.  They
are often filled with technical details uncommon for that part of the
scientific audience which is not directly involved in this research direction.
That is why this is so important to have simple and intuitively clear
interpretations of the relativistic calculations, and an explicit
systematic method to compare the relativistic and nonrelativistic
results.

In the present work we are going to
analyze the extended calculations of the static
properties of the deuteron utilizing the Bethe-Salpeter (BS) amplitudes
which are recently computed numerically \cite{umnikov-khanna}.
The main goal of
our paper is to contribute to the development of the physical
intuition for understanding the relativistic calculations and their
comparison to the nonrelativistic calculations.  Our basic idea is to
compute the observable densities of various charges (e.g., vector and
axial-vector charges) in both the relativistic and the
nonrelativistic formalisms
and use these densities as tools to compare relativistic amplitudes
and nonrelativistic wave functions, which can not be rigorously
interrelated otherwise.
In doing so we pursue, in some sense, the goals
opposite to the ones we outlined above as typical for the approach within
the covariant description of the deuteron.  Another goal of our paper is
to fill some gap in the literature by giving explicit expressions
relating the BS amplitudes in different representations,
which will help to compare the relativistic amplitudes computed in
different models.

We calculate here the magnetic and quadrupole moments of the
deuteron within the Bethe-Salpeter formalism. The investigation of these
static characteristics of the deuteron is still an important
topic in nuclear physics. In the nonrelativistic models it gives the
direct information about the tensor components in the NN interaction
and the magnitude of the $D$ wave probability in the deuteron.
However, there is an essential problem in fitting the experimental
values of the quadrupole and magnetic moments within the same $D$ wave
probability in the nonrelativistic calculations
(cf.\ ref.\ \cite{lommon} and references therein).
The efforts, aiming to solve this difficulty, go in two main directions,
namely calculating the corrections of the meson exchange
currents \cite{tamura,gary,mec}
and taking into account the relativistic
effects \cite{tjon,gross,ZT,RT,gross2,honzawa,brodsky}.  In the conventional
approach   the mesonic degrees of freedom and relativistic
effects are treated as corrections to the nonrelativistic potential theory.
It is found that, by adding these effects to the quadrupole moment,
a satisfactory description of the data may be achieved for
a broad range of different potentials \cite{lommon},
while the magnetic moment
shows an essential sensitivity to the model calculations of
the meson exchange currents. Moreover, the consistency
of such calculations is not at all clear. For this reason
a comprehensive covariant investigation has its own right.
A prominent feature of the
relativistic consideration within the Bethe-Salpeter formalism is
that the meson exchange effects due to pair creation currents is
taken into account consistently \cite{lommon,karmanov,dorkin}, so that
the essential part of the mentioned effects may be estimated in a
self consistent way.

The general approach to calculate  the
static characteristic of the deuteron within the
BS formalism has been elaborated by several authors since some time
(see, for instance,  refs.\ \cite{gross2,ZT,RT,honzawa,bondaren}) and
numerical estimates have been performed. However, explicit calculations
have been done within additional approximations for the
solution of the BS equation, that is,
for a separable interaction and by disregarding the negative
energy states \cite{RT}, or with one nucleon on mass-shell \cite{gross2},
or from a general point of view with adjusting the probability of
the $P$ states as to fit simultaneously both the quadrupole and
magnetic moments \cite{honzawa} (for this goal one needs an anomalously
large pseudoprobability of the $P$ waves, say $\sim 1.5\%$).
In this paper we perform a covariant calculation of the quadrupole
and magnetic moments of the deuteron within the exact solution of the
BS equaion and avoid additional approximations to the problem.

Our present investigation is also
partially motivated by the renewed interest in the experimental
investigation of the nucleon and deuteron spin-dependent structure
functions at low
momentum transfer $Q^2$ \cite{cebaf}.  This interest is connected to
the study of the $Q^2$-evolution of the Gerasimov-Drell-Hearn sum
rule \cite{gdh}, which relates the spin-dependent structure functions
of the targets to their the magnetic moments.  For instance, only a correct
description of the deuteron
magnetic and quadrupole moments will assure a reliable extraction
of the information about the neutron structure function   from
the deuteron data.

Our paper is organized as follows.
In Section II  the basic covariant formulae for the
electromagnetic current and static moments of the deuteron are presented.
In Section III  the general definitions
of the Bethe-Salpeter amplitudes for the deuteron
are given in different representations and their symmetry properties
are studied in detail. The transformation matrix
relating the amplitudes in different representations
is determined. The relativistic amplitudes are compared
to the nonrelativistic wave functions, using the
calculated observables, e.g., the vector and axial charge densities.
In Section IV  the covariant formulae for the magnetic and quadrupole
moments are derived  in the Breit frame.
The effects of the Lorentz deformation
and dependence of the amplitude on the relative energy of the
two nucleons in the deuteron are explicitly taken into account.
The terms corresponding to the nonrelativistic
expressions for the moments are determined in explicit form and the
relativistic corrections are computed.
The Sections V and VI contain conclusions and the summary, respectively.

\section{Relativistic kinematics of the electromagnetic current}

The definition of the quadrupole moment $Q_D$
and the magnetic moment $\mu_D$ of the deuteron
appears most transparent if one starts with the famous
Rosenbluth formula for the elastic scattering of electrons off the deuteron,
$e+D\to e^\prime+D^\prime$,
\begin{eqnarray}
\frac{d\sigma}{d\Omega}\biggl|_{Lab}
=\frac{d\sigma}{d\Omega}\biggl|_{Mott}
\left(A(q^2)+B(q^2)\tan^2\frac{\theta}{2}\right)\label{rosen}
\end{eqnarray}
with the following decomposition of the electromagnetic form factors
\begin{eqnarray}
&&
A(q^2)=F_{C}^2(q^2)+{8\over 9}
\eta^2F_Q^2(q^2)+{2\over 3}\eta F_{M}^2(q^2),\label{formfA}\\
&&B(q^2)={4\over 3}\eta(1+\eta)F_M^2(q^2),\label{formfB}
\end{eqnarray}
where $\eta=-{q^2}/{4M_D^2}$ and $M_D$ is the deuteron mass.
$Q^2 = - q^2$ denotes the momentum transfer.
Then the quadrupole and magnetic moments of the
deuteron are defined via the normalization conditions for
the charge ($F_{C}$), quadrupole ($F_Q$) and magnetic ($F_{M}$)
form factors at vanishing momentum transfer $q^2=0$
\begin{eqnarray}
F_{C}(0)=1,
\quad
F_Q(0)=M_D^2 \, Q_D,
\quad
F_{M}(0)=\mu_D\frac{M_D}{m}
\label{norm}
\end{eqnarray}
with $m$ as nucleon mass.

The general form of the deuteron electromagnetic current
which is invariant under
the Lorentz and time-reverse transformations, is given by
\begin{eqnarray}
\langle P^\prime, \lambda^\prime |J_\mu|P,\lambda\rangle=
-\frac{e}{2M_D}\varepsilon^*_\rho({\bf P}^\prime,
\lambda^\prime)J_\mu^{\rho\sigma}
\varepsilon_\sigma({\bf P},\lambda),\label{current}
\end{eqnarray}
where $\varepsilon^{*}({\bf P}^\prime,\lambda^\prime)$ and
$\varepsilon({\bf P},\lambda)$ are the polarization four vectors
of the initial and final deuteron states. The
covariant normalization  of the current reads
\begin{eqnarray}
\lim_{q^2 \to 0}
\langle P^\prime,
\lambda^\prime |J_\mu|P,\lambda \rangle =
e \, \frac{P_\mu}{M_D} \, \delta_{\lambda^\prime,\lambda}.
\label{normalization}
\end{eqnarray}
The matrix element
$J^\mu_{\rho\sigma}$
can be expanded in terms of the scalar form factors in the form
\begin{eqnarray}
J^\mu_{\rho\sigma}=(P^\prime+P)^\mu\left(
g_{\rho\sigma}F_1(q^2)-\frac{q_\rho q_\sigma}{2M_D^2}F_2(q^2)\right)
+(g_\rho^\mu q_\sigma - g_\sigma^\mu q_\rho)G_1(q^2).\label{oper}
\end{eqnarray}
The scalar form factors $F_{1,2}$
and $G_1$ are related to the  form factors $F_{C,Q,M}$,
by (cf.\ \cite{gourdin})
\begin{eqnarray}
&&F_{C}(q^2)=
F_1(q^2)+{2\over 3}\eta
[F_1(q^2)+(1+\eta)F_2(q^2)-G_1(q^2)], \label{charge}\\
&&F_Q(q^2)=F_1(q^2)+(1+\eta)F_2(q^2)-G_1(q^2), \label{quad}\\
&&F_{M}(q^2)=G_1(q^2). \label{mag}
\end{eqnarray}

In the nonrelativistic impulse approximation
these deuteron form factors read
\begin{eqnarray}
&&F_{C}(q^2)=(G_E^p(q^2)+G_E^n(q^2))\, C_E(q^2),\\
&&F_Q(q^2)=(G_E^p(q^2)+G_E^n(q^2))\, C_Q(q^2),\\
&&F_{M}(q^2)=\frac{M_D}{m}\left[
(G_E^p(q^2)+G_E^n(q^2))\, C_S(q^2)+\frac12(G_M^p(q^2)+G_M^n(q^2))
\, C_L(q^2)\right],
\end{eqnarray}
where $G_E^{(p,n)}(q^2)\, \left (G_M^{(p,n)}(q^2)\right )$
are the electric (magnetic) nucleon formfactors and
the invariant functions $C(q^2)$ are defined  by
\begin{eqnarray}
&&C_E(q^2)=\int\limits_0^\infty (u^2+w^2) \,
j_0\left(\frac{qr}{2}\right)dr,\ \ C_E(0)=1,
\label{c1}\\[2mm]
&&C_Q(q^2)= \frac{3\sqrt{2}}{2\eta}\int\limits_0^\infty
\left(uw-\frac{w^2}{2\sqrt{2}}\right)
j_2\left(\frac{qr}{2}\right)dr,\ \ C_Q(0)=M_D^2 \, Q_D,
\label{c2}\\[2mm]
&&C_L(q^2)=\frac32\int\limits_0^\infty w^2
\left[j_0\left(\frac{qr}{2}\right) +
j_2\left(\frac{qr}{2}\right)
\right]dr,\ \ C_L(0)=\frac32 P_D,
\label{c3}\\[2mm]
&&C_S(q^2)=\int\limits_0^\infty \left(u^2-\frac{w^2}{2}\right)
j_0\left(\frac{qr}{2}\right)dr
+\int\limits_0^\infty \left(\frac{uw}{\sqrt{2}}+\frac{w^2}{2}\right)
j_2\left(\frac{qr}{2}\right)dr,
\nonumber\\
&& \hskip 10cm C_S(0)=1-\frac32 P_D.
\label{c4}
\end{eqnarray}
Here $j_i$ is the modified Bessel function of $i$-th order,
$u$ and $w$ represent the S and D waves of the nonrelativistic
deuteron wave function, and $P_D$ is the weight of the D wave
in the deuteron wave function.

To calculate the form factors $F_{C,Q,M}$ within the
Bethe-Salpeter formalism one has to express the current
(\ref{current}) in terms of the BS amplitudes and, then,
to extract the coefficients of different Lorentz structures
given by eq.~(\ref{oper}). Taking the limit $q^2\to 0$,
the static moments can be also obtained. Apparently,
these calculations can be done in any particular reference frame.
For example, the Breit frame is
especially convenient for such type of calculations.
The  Breit frame for the deuteron is defined by
\begin{eqnarray}
{\bf P}=-\frac{\bf q}2,\quad {\bf P}^\prime=\frac{\bf q}2,\quad
P_0=P_0^\prime=E.
\end{eqnarray}

Choosing ${\bf q}$ along the positive $z$ axis and
contracting
$J^\mu_{\rho\sigma}$ and the polarization vectors, which obey
$\varepsilon({\bf P}^\prime,\lambda^\prime)$,
 $\varepsilon({\bf P},\lambda)$:
\begin{eqnarray}
&&\varepsilon({\bf P}^\prime,1)=\varepsilon({\bf P},1)=
-{1\over \sqrt{2}}(0,1, i ,0),\\
&&\varepsilon({\bf P}^\prime,-1)=\varepsilon({\bf P},-1)=
{1\over \sqrt{2}}(0,1,- i ,0),\\
&&\varepsilon({\bf P},0)=(-\sqrt{\eta},0,0,\sqrt{1+\eta}),\quad
\varepsilon({\bf P}^\prime,0)=(\sqrt{\eta},0,0,\sqrt{1+\eta}),
\end{eqnarray}
one arrives at the expressions
for the matrix elements of the deuteron electromagnetic current
in terms of the form factors:
\begin{eqnarray}
&&\langle P^\prime,\lambda^\prime |J^0|P,\lambda\rangle=
e\sqrt{1+\eta}\left(
F_1\delta_{\lambda\lambda^\prime}+2\eta\left[
F_1+(1+\eta)F_2-G_1\right]
\delta_{\lambda^\prime, 0}\delta_{\lambda, 0}\right),\label{moi1}
\\[2mm]
&&\langle P^\prime,\lambda^\prime |J^x|P,\lambda\rangle=
e\frac{\sqrt{\eta}}{2}\sqrt{1+\eta}G_1
\left(\delta_{\lambda^\prime, \lambda+1} -
\delta_{\lambda^\prime, \lambda-1}\right),\label{moi2}\\[2mm]
&&\langle P^\prime,\lambda^\prime |J^y|P,\lambda\rangle=- i
e\frac{\sqrt{\eta}}{2}\sqrt{1+\eta}G_1
\left(\delta_{\lambda^\prime, \lambda+1} +
\delta_{\lambda^\prime, \lambda-1}\right),\label{moi3}\\[2mm]
&&\langle P^\prime,\lambda^\prime |J^z|P,\lambda\rangle=0.
\label{moi4}
\end{eqnarray}
Thus the magnetic and quadrupole form factors of the deuteron
are recovered by
\begin{eqnarray}
&&\mu_D=\frac{m}{M_D}\sqrt{2}\lim_{\eta\to 0}
\frac{\langle P^\prime,\lambda^\prime=1 |J^x|P,
\lambda=0\rangle}{\sqrt{\eta}\sqrt{1+\eta}},
\label{magnetic}\\[2mm]
&&Q_D=\frac{1}{M_D^2}\lim_{\eta\to 0}
\frac{\langle P^\prime,\lambda^\prime=0 |J^0|P,
\lambda=0\rangle-\langle P^\prime,\lambda^\prime=1
 |J^0|P,\lambda=1\rangle}{2\eta\sqrt{1+\eta}}.
\label{quadrupole}
\end{eqnarray}
Equations (\ref{moi1})-(\ref{quadrupole}) are the basic
relations providing the
calculations of the electromagnetic characteristics of the deuteron.
In practice, one needs to define explicitly the operator $J_\mu$ of the
electromagnetic current and calculate its matrix
elements with the deuteron states
$|P,\,\lambda\rangle$.

\section{The bound state wave function}

\subsection{General definitions}

Using the technique presented in ref.\ \cite{umnikov-khanna},
the BS equation  for a bound state in
ladder approximation can be written in the form
\begin{eqnarray}
&&K(p_0,{\bf p}) \, \chi(p;P)+
\sum\limits_{B}
\frac{\lambda_B}{4 i \pi^3}\int\!d^4p^\prime
\frac{\Lambda(p_1) \, \Gamma_B \, \chi(p^\prime;P) \,
\Gamma_B \, \Lambda(p_2)}
{(p-p^\prime)^2-\mu^2_B} = 0,\label{eqn-to-solve}\\
&&K(p_0,{\bf p})\equiv{\left (E_{\bf p}^2-p_0^2-M_D^2/4\right)^2
-p_0^2M_D^2},
\label{prop}
\end{eqnarray}
where $\chi(p;P)$ is the BS amplitude for the deuteron in the
matrix representation \cite{umnikov-khanna};
$\Lambda(p_i)=\hat{p_i}-m$;  $p=(p_0,{\bf p})$
is four-momentum of the $i$-th nucleon expressed in terms
of relative four-momenta $p$ or $p^\prime$
and the center-of-mass (c.m.) momentum $P=(M_D,{\bf 0})$:
$p_{1,2} = P/2 \pm p$;
$B$ enumerates the exchanged mesons
$\pi, \sigma, \omega, \rho, \delta, \eta$;
$\mu_B$ is the mass of the meson;
 $\Gamma_B$ is the interaction vertex between the nucleon
and the corresponding boson $B$ and
$\lambda_B \equiv g_B^2/4\pi$ with $g_B$ being the coupling constant.
We use here the short hand notation $\hat p \equiv p^\mu \, \gamma_\mu$.

Since the BS amplitude $\chi$ and its adjoint $\bar\chi$
satisfy the  homogeneous BS equation they are
determined up to an arbitrary constant
which is fixed by  additional normalization condition.
In the ladder approximation
the normalization constant may be fixed by computing
the matrix element of the electromagnetic
current at $q^2=0$
\begin{eqnarray}
\int\!\frac{d^4 p}{ i (2\pi)^4} \,
{\mbox Tr}\left\{\bar\chi(p;P) \, \gamma_\mu\,
\chi(p;P)(m-\hat p_1)\right\}=2P_\mu.
\label{ladder-norm}
\end{eqnarray}
The normalization condition
(\ref{ladder-norm}) coincides with  the one used in  ref.\ \cite{ZT}.

The BS amplitude is a $(4\times 4)$ matrix in the spinor space,
and consequently the
BS equation (\ref{eqn-to-solve}) possesses this matrix structure as well.
To solve this matrix equations one can utilize a decomposition of
the BS amplitude over
a complete set of $(4\times 4)$ matrices and solves a system of coupled
equations for the coefficients of such a decomposition. The
choice of the representation
of the matrices depends on the concrete attacked problem.
Certainly,
different representations are related by linear transformations,
and it is straightforward (but cumbersome)
to transform results from one  representation to  another one.
In our opinion,
to solve the BS equation and to compute matrix elements of the deuteron
observables (as for instance eq.~(\ref{ladder-norm}))
a convenient way is to decompose the amplitude in terms
of the complete set of Dirac matrices, which form the Clifford algebra
(for more details cf.\ ref.\ \cite{umnikov-khanna}).
By exploiting the parity invariance of the BS amplitude
\begin{eqnarray}
{\cal P} \chi_D(p_0, {\bf p}) = \eta_{{\cal P}}
\gamma_0 \chi_D(p_0, -{\bf p}) \gamma_0,
\label{parity}
\end{eqnarray}
it may be written for the deuteron, which has positive parity eigenvalues
$(\eta_{{\cal P}}=1$), as
\begin{eqnarray}
\chi_D(p;P)=\gamma_5{\sf P}
+\gamma^5\gamma^0{\sf A}^0
-(\bfgr{\gamma}\!\cdot\!{\vec {\sf V}})-
\gamma_5(\bfgr{\gamma}\!\cdot\!{\vec {\sf A}})
-2 i \gamma^0(\bfgr{\gamma}\!\cdot\!{\vec {\sf T}}^0) -
2\gamma^0\gamma_5(\bfgr{\gamma}\!\cdot\!{\vec {\sf T}}),
\label{sys-deuteron}
\end{eqnarray}
with pseudoscalar (${\sf P}$, ${\sf A}^0$),
axial ($\vec {\sf A}$)
and vector (${\vec {\sf T}}^{0}$, ${\vec {\sf T}}$, ${\vec {\sf V}}$)
functions depending only upon the relative four momentum $p$ in the
c.m. frame.
The angular dependence of the state with
spin  $J=1$ and  its projection
${\cal M}$ owing
to the rotational invariance of eq.~(\ref{eqn-to-solve}) is expressed in
terms of the spherical and vector spherical harmonics. For example, when
denoting
${\sf X} = ({\sf P}, {\sf A}^0)$ and
$\vec {\sf X} = (\vec {\sf A}, {\vec {\sf T}}^{0}, {\vec {\sf T}},
{\vec {\sf V}})$, we may write
\begin{eqnarray}
{\sf X}(p_0,{\bf p}) =
{\sf X}_1(p_0,|{\bf p}|)
{\mbox{\large {Y}}}_{1{\cal M}}(\Omega_p),\quad
{\vec {\sf X}}(p_0,{\bf p})=
\sum_{L=0,1,2}
{\sf X}_{L}(p_0,|{\bf p}|)
{\mbox{\large \bf {Y}}}_{1{\cal M}}^{L}(\Omega_p). \label{har}
\end{eqnarray}
The corresponding equations for the radial functions can be found
by a partial wave decomposition of the kernel in eq.~(\ref{eqn-to-solve})
and by carrying out the angular
integration. An example of the system of coupled equation
for the radial amplitudes in the case of one scalar exchanged
boson is given in the ref.\ \cite{umnikov-khanna}.
In what follows  the notation for the radial
amplitudes are kept as in eq.~(\ref{sys-deuteron})
with the lower index
indicating the value of the angular momentum $L$ in eq.~(\ref{har}).

\subsection{The transformation properties of the partial amplitudes}

Due to the parity invariance, (\ref{parity}),
only eight radial components are
relevant to describe the deuteron amplitude, namely
\begin{eqnarray}
{\sf P}_1, \quad{\sf A}^0_1, \quad{\sf A}_0, \quad{\sf A}_2, \quad
{\sf V}_1, \quad{\sf T}^0_1, \quad{\sf T}_0, \quad{\sf T}_2.
\label{sys2}
\end{eqnarray}
Analyzing the behavior of the amplitude under the symmetry transformations,
one can establish the properties of the components (\ref{sys2}).
The invariance of the BS equation under the time-reversal
operation ${\cal T}$
\begin{eqnarray}
{\cal T} \chi^D_{{\cal M}}(p_0, {\bf p}) =
\gamma^1 \, \gamma^3 \, {\chi^D_{\cal M}}^\ast(p_0, -{\bf p})
\gamma^1 \, \gamma^3
\label{time}
\end{eqnarray}
and the complex conjugation  ${\cal K}$
\begin{eqnarray}
{\cal K}{\chi^D_{{\cal M}}}(p) = {(-1)}^{{\cal M}}\chi^D_{-{\cal M}}(p)
\label{complex}
\end{eqnarray}
imply that the seven partial amplitudes
${\sf P}_1$, ${\sf V}_1$, ${\sf A}^0_1$, ${\sf A}_{0,2}$,
${\sf T}_{0,2}$  are real functions while the amplitude
${\sf T}_1^0$  is purely imaginary, i.e.,
${{\sf T}_1^0}^\ast = - {\sf T}_1^0$.

The Pauli principle  implies that the amplitude $\chi_D(p)$
changes the sign if two nucleons are interchanged, i.e.,
\begin{eqnarray}
\chi_D(p_0, {\bf p}) = - \chi_D^+(-p_0, - {\bf p}).
\label{pauli}
\end{eqnarray}
>From eqs.~(\ref{pauli}) and (\ref{complex}) follows that
${\sf A}_1^0$ and  ${\sf T}_1^0$ are odd functions with respect
to the operation $\Pi(p_0\to - p_0)$
\begin{eqnarray}
\Pi \, {\sf A}_1^0(p_0,{\bf p}) = - {\sf A}_1^0(p_0, {\bf p}),
\quad
\Pi \, {\sf T}_1^0(p_0,{\bf p}) = - {\sf T}_1^0(p_0, {\bf p})
\label{otnos}
\end{eqnarray}
and the remaining six amplitudes are even functions of $p_0$.
This symmetry property is
useful for the classification of the amplitude according to two-nucleon
states with a given relative energy, i.e.,
the $\rho$ spin classification.

Table~\ref{ampBS} summarizes the  properties of the
partial BS amplitudes in the representation
(\ref{sys-deuteron}) under the symmetry transformations.

\subsection{Observables}

Relying on the symmetry properties of the partial amplitudes,
defined by eq.~(\ref{sys-deuteron}), the BS
equation (\ref{eqn-to-solve})
has been solved numerically \cite{umnikov-khanna} for the deuteron at rest
by performing a Wick rotation ($p_0\to ip_4$).
In our present calculations we include six meson exchanges,
$\pi$, $\omega$, $\rho$, $\sigma$, $\eta$ and $\delta$,
which describe the effective $NN$-forces.
The set of the meson parameters, such as masses, coupling constants
and cut-off form factors, has been taken essentially the same
as in ref.\ \cite{ZT}, where it has been obtained from a fit of
the phase-shift analysis of the $NN$ scattering and
the binding energy of the deuteron.

The BS amplitude does not have a direct probabilistic
interpretation as the Schr\"odinger  wave function.
Moreover,
there is no simple way to compare these two objects
describing the same system, the deuteron.
In order to make a comparison possible, we
can compute the same matrix elements of observables
in two approaches and compare these observables.

For example, the normalization
condition (\ref{ladder-norm}) in the rest frame of the
deuteron and for $\mu = 0$,
$\langle D|\bar N(0)\gamma^0 N(0)|D\rangle=2M_D$, is
simply a charge of the deuteron associated with vector current.
In the Wick rotated system
 ($p_0 \to i p_4$) and in terms of the partial
 amplitudes~(\ref{sys2}) it reads
\begin{eqnarray}
M_D& = & 2\int\!\frac{d{p_4} \, d|{\bf p}| |{\bf p}|^2}{(2\pi)^4}
\Bigl\{-M_D\left({\sf P}_1^2 + {{\sf A}^0_1}^2 + 4{{\sf T}^0_1}^2
+ {\sf V}_1^2\right) \nonumber \\[2mm]
& + &
(2m_N-M_D)\left({X_0^+}^2+{X_2^+}^2\right)-(2m_N+M_D)
\left({X_0^-}^2+{X_2^-}^2\right) \nonumber \\[2mm]
& + &
\frac{2\sqrt{2}|{\bf p}|}{\sqrt{3}}{\sf P}_1
\left({\sf X}_0^+ - \sqrt{2}{\sf X}_2^+ + {\sf X}_0^-
- \sqrt{2}{\sf X}_2^-\right)
\nonumber \\
& - &
\frac{2\sqrt{2}|{\bf p}|}{\sqrt{3}}{\sf V}_1
\left(\sqrt{2}{\sf X}_0^+ + {\sf X}_2^+
- \sqrt{2}{\sf X}_0^- - {\sf X}_2^-\right)
\Bigr\},
\label{normdc}
\end{eqnarray}
where
${\vec{\sf X}}^\pm\equiv\sqrt{2}({\vec{\sf T}}\pm{\vec{\sf A}}/2)$.
Now we
define the charge density $\rho_{\rm ch}(|{\bf p}|)$ as
\begin{eqnarray}
&&\frac{1}{2M_D}
\langle D|\bar N(0) \gamma^0 N(0)|D\rangle =
\int\!\frac{d{p_4} \, d|{\bf p}||{\bf p}|^2}{(2\pi)^4}
\rho_{\rm ch}({p_4}, |{\bf p}|),
\label{nr1}\\[2mm]&&
\rho_{\rm ch}(|{\bf p}|) \equiv
\int\limits_{-\infty}^{\infty}\!\frac{d{p_4}}{2\pi} \, \rho_{\rm ch}
({p_4}, |{\bf p}|).
\label{charged}
\end{eqnarray}
This already may be
compared with the corresponding nonrelativistic analogue, i.e.,
the square of the deuteron wave function in the momentum space,
$\propto u^2(p) +w^2(p)$.

In the same manner also the nucleon spin-density may be defined
as density of the axial charge
\begin{eqnarray}
&&
\frac{1}{2M_D}
\langle D|\bar N(0) \gamma_5\gamma^0 N(0)|D\rangle =
\int\!\frac{d{p_4} d|{\bf p}||{\bf p}|^2}{(2\pi)^4}
\rho_{\rm spin}({p_4}, |{\bf p}|),
\label{nr2}\\[2mm]&&
\rho_{\rm spin}(|{\bf p}|) \equiv
\int\limits_{-\infty}^{\infty}\!\frac{d{p_4}}{2\pi}\rho_{\rm spin}
({p_4}, |{\bf p}|).\label{spin}
\end{eqnarray}
In the nonrelativistic
limit this density
reflects the contribution of the $D$-wave admixture in the deuteron,
$\propto u^2(p) - \frac12 w^2(p)$.

Results of numerical calculation of the defined
densities together with a comparison with their nonrelativistic
counterparts obtained within the Bonn and Paris potentials
are presented in figs.~\ref{chargedens} and~\ref{spindens}.
All curves exhibit qualitatively similar shapes and are identical
in the nonrelativistic region $|{\bf p}| \le 0.5$ GeV/c.
If the momentum $|{\bf p}| $ increases, the deviations of the relativistic
results from the nonrelativistic ones becomes more significant, but still
too small to be attributed to the relativistic
effect. Rather it is compatible with the model differences.
Particular attention is to be paid to the fig.~\ref{spindens}, where
the spin density is depicted. This function is rather sensitive
to the internal spin-orbital structure of the deuteron.
The fact
that the "elementary oscillations" of the spin density in
the potential models
are reproduced by the solution of the BS equation  might be interpreted
as the relativistic structure of the deuteron is
governed by the nucleon interaction in states with a positive energy
and $L=0,\,2$, i.e., by $^3S_1$ and $^3D_1$ configurations.
Therefore, in spite of the
quadratic forms of the partial amplitudes is  not diagonal
in eqs.~(\ref{normdc}) and  (\ref{spin}), one can define the
relativistic analog of the probability of the
$D$-wave admixture in the deuteron.
Carrying out the $|p|$ integration in eq.~(\ref{spin})
and equating the result to $(1-3/2\,P_D)$ we find $P_D\approx 5\%$
(cf.\ \cite{ZT}), which is compatible   with the probabilities of the
Bonn ($P_D = 4.3$ \cite{bonn}) and Paris ($P_D = 5.9$ \cite{paris}) potential
models.

\subsection{The BS amplitude in different representations}

To have a closer analogue with the nonrelativistic
consideration it is convenient to
use another basis set of matrices in the decomposition of
the BS amplitude.
In the literature the two-spinor basis
\cite{cubis} is frequently used, which means
an outer product of two spinors, representing solutions of the
free Dirac equation with positive and negative energies.
This basis is labeled by the relative momentum $\vec p$, the helicities
$\lambda_i$ and the energy spin $\rho_i$ of the particles \cite{ZT},
sometimes also called
$(J,\lambda_1,\lambda_2,\rho_1,\rho_2)$ representation. In this case
one usually adopts for the partial amplitudes
the spectroscopic notation $^{2S+1}L^{\rho_1,\rho_2}_J$, i.e.,
\begin{equation}
^3S^{++}_1, \,^3S^{--}_1,\,
^3D^{++}_1,\, ^3D^{--}_1, \, ^1P^{+-}_1,
\, ^1P^{-+}_1,\, ^3P^{+-}_1, \,^3P^{-+}_1.
\label{spectr}
\end{equation}
Sometimes it is more convenient to change from the
$(J,\lambda_1,\lambda_2,\rho_1,\rho_2)$ representation to
the representation $(J,L,S,\rho$)
where $\rho$ is the projection of the total energy spin
of the system. In this case the notation of the components is as follows
\begin{eqnarray}
Y^T\equiv (v_s^o, v_t^e, v_s^e, v_t^o, u^+, u^-, w^+, w^-),
\label{your}\end{eqnarray}
where $u,v,w$ correspond to $L=0,1,2$ respectively and $o$ or $e$
mean the odd or even parity relative to the $\rho$-spin
function;
the lower indices $s,t$ denote the singlet and triplet
spin configurations respectively.
According to eqs.~(\ref{pauli}) and (\ref{complex}), the amplitudes
$v_s^o, v_t^e$ are odd  and $v_s^e, v_t^o$ are even functions of $p_0$.
The partial amplitudes in the basis (\ref{spectr}), (\ref{your})
are of a more familiar form and
show a more transparent physical meaning since they may
be compared with the deuteron states in the nonrelativistic limit.
It is intuitively clear
(see also figs.~\ref{chargedens} and~\ref{spindens})
that the two nucleons in the deuteron are mainly in
states with $L=0,2$ and with positive energy so that one may
expect that the probability of
states with negative energies and $L=1$ in
eqs.~(\ref{spectr}) - (\ref{your}) is much smaller in comparison
with the probability for the
$^3S^{++}_1$ and $^3D^{++}_1$ (or $  u^+$  and  $w^+$)
configurations.
Moreover, it can be shown that the waves
$^3S^{++}_1$ and $^3D^{++}_1$  directly correspond to the
$S$ and $D$ waves in the deuteron, while
those with the negative energy vanish
in the nonrelativistic limit.

The partial amplitudes (\ref{spectr})
are defined through the following decomposition
of the BS amplitude
\begin{equation}
\chi_D(p_0, {\bf p}) = \sum\limits_\alpha
\phi_\alpha (p_0, |{\bf p}|){\cal V}^\alpha_{\cal M}({\bf p}),
\label{razlozhenie}
\end{equation}
where $\alpha = \{ J, L, S, \rho_1, \rho_2 \} $
labels different states of the system;
$\phi_\alpha$ denotes the partial
amplitudes  in eq.~(\ref{your}),
${\cal V}^\alpha_{\cal M}({\bf p})$
are the spin-angular functions
\begin{equation}
{\cal V}^\alpha_{\cal M}({\bf p})=i^L
\sum\limits_{s_1 s_2 m}(LmSs|J{\cal M})(\frac{1}{2}s_1
\frac{1}{2}s_2|Ss) Y_{Lm}(\hat {\bf p}) U_{s_1}^{\rho_1}({\bf p})
U_{s_2}^{\rho_2}(-{\bf p}).
\label{spinangle}
\end{equation}
In eq.~(\ref{spinangle}) the quantities $U_{s}^{\rho}({\bf p})$ are the
free nucleon spinors; the explicit matrix form
for the spin-angular functions
${\cal V}^\alpha_{\cal M}({\bf p})$
is given in the Appendix I.

In order to establish a connection between the representation
(\ref{sys-deuteron}) and the spinor basis
(\ref{spectr})-(\ref{razlozhenie})
we represent the Dirac
matrices in eq.~(\ref{sys-deuteron})
as a direct product of Pauli matrices of the nucleon
spin $\bfgr{\sigma}$ and the $\rho$-spin
\begin{eqnarray}       &&
\chi_D(p)=
\rho^1\!\otimes\!
[\hat{ I} {\sf P}-2i(\bfgr{\sigma}\!\cdot\!{\vec {\sf T}}_0)]
+i\rho^2\!\otimes\![\hat{ I}{\sf A}^0 +
(\bfgr{\sigma}\!\cdot\!{\vec {\sf V}})] +
\rho^3\!\otimes\!(\bfgr{\sigma}\!\cdot\!{\vec {\sf A}})
+2I\!\otimes\!(\bfgr{\sigma}\!\cdot\!{\vec {\sf T}}).
\label{sys-rho}\end{eqnarray}
The last two terms in eq.~(\ref{sys-rho}) may be rewritten as
\begin{eqnarray}
&&\rho^3\!\otimes\!(\bfgr{\sigma}\!\cdot\!{\vec {\sf A}})
+2I\!\otimes\!(\bfgr{\sigma}\!\cdot\!{\vec {\sf T}}) =
\frac{1}{\sqrt{2}}(\hat{I}+\rho^3)\!\otimes\!(\bfgr{\sigma}\!\cdot\!{\vec{\sf X}}^+)
+\frac{1}{\sqrt{2}}(\hat{I}-\rho^3)\!\otimes\!(\bfgr{\sigma}\!\cdot\!{\vec{\sf X}}^-),
\label{AB}\end{eqnarray}
Then eqs.~(\ref{sys-rho}) and~(\ref{AB}), together with the symmetry
properties of our partial amplitudes listed in the Table~\ref{ampBS},
show that the desired relation between the two representations
appears as follows:
\begin{eqnarray*}
&&
^3S_1^{++} \sim {\sf X}_0^+,\,\, ^3S_1^{--} \sim {\sf X}_0^-,
\,\,
^3D_1^{++} \sim {\sf X}_2^+,\,\,
^3D_1^{--} \sim {\sf X}_2^-,\\[2mm]
&&
^3P_1^{e} \sim {\sf T}_1^0,\,\,   ^3P_1^{o} \sim {\sf V}_1,\,\,
^1P_1^{e} \sim {\sf P}_1 \,\,   ^1P_1^{o} \sim {\sf A}_1^0 .
\end{eqnarray*}
The relation between (\ref{sys2}) and (\ref{your})
can be established exactly.
The components being odd in the relative energy
$v_s^o$, $v_t^e$ and ${\sf A}_0^1$, ${\sf T}_0^1$ are related directly
to each other via
\begin{eqnarray}
v_s^o=-i{\sf A}_0^1,\quad v_t^e=2{\sf T}_0^1,
\label{odd}
\end{eqnarray}
whereas the remaining six components are connected via linear combinations.
By representing these amplitudes as six-component vectors,
$\tilde Y^T = (v_s^e, v_t^o, u^+, u^-, w^+, w^-)$ and
$\Psi^T=({\sf P}_1,\, {\sf V}_1, \, {\sf X}_0^+, \, {\sf X}_0^-,\,
{\sf X}_2^+,\,{\sf X}_2^-)$, the transition from
$\tilde Y$ to $\Psi$ is provided by a unitary transformation
$\tilde Y = U \Psi$ (with det$(U)=-1$, and $U\,U^T=1$) with the
following explicit form of the transition matrix
\begin{eqnarray}
U&\!\!\!=\!\!\!&\frac{\zeta}{2\sqrt{1+\zeta^2}}\label{U-mat}\\[1mm]
&\!\!\!\times\!\!\!&\left(
\begin{array}{cccccc}
-\frac{2}{\zeta} & 0 & \sqrt{{2\over 3}} & 
- \sqrt{{2\over 3}} & - \frac{2}{\sqrt{3}} & \frac{2}{\sqrt{3}} \\[2mm]
 0 & \frac{2}{\zeta} & \frac{2}{\sqrt{3}} &
  \frac{2}{\sqrt{3}} & \sqrt{{2\over 3}} & \sqrt{{2\over 3}} \\[2mm]
\sqrt{{2\over 3}} & -\frac{2}{\sqrt{3}} & 
\frac{1+\sqrt{1+\zeta^2}}{\zeta} & \frac{1-\sqrt{1+\zeta^2}}{3\zeta} 
& 0 & \frac{2\sqrt{2}}{3}\frac{1-\sqrt{1+\zeta^2}}{\zeta} \\[2mm]
-\sqrt{{2\over 3}} & -\frac{2}{\sqrt{3}} & 
\frac{1-\sqrt{1+\zeta^2}}{3\zeta} & \frac{1+\sqrt{1+\zeta^2}}{\zeta}
 & \frac{2\sqrt{2}}{3}\frac{1-\sqrt{1+\zeta^2}}{\zeta} & 0 \\[2mm]
\frac{2}{\sqrt{3}} & \sqrt{{2\over 3}} & 0 & 
- \frac{2\sqrt{2}}{3}\frac{1-\sqrt{1+\zeta^2}}{\zeta}
 & -\frac{1+\sqrt{1+\zeta^2}}{\zeta} & \frac{1-\sqrt{1+\zeta^2}}{3\zeta} \\[2mm]
-\frac{2}{\sqrt{3}} & \sqrt{{2\over 3}} & - 
\frac{2\sqrt{2}}{3}\frac{1-\sqrt{1+\zeta^2}}{\zeta} & 
0 & \frac{1-\sqrt{1+\zeta^2}}{3\zeta} & -\frac{1+\sqrt{1+\zeta^2}}{\zeta}
\end{array}\right), \nonumber
\end{eqnarray}
with $\zeta = |{\bf p}|/m$.
In the nonrelativistic limit, where $\zeta\!\ll\!1$, the
matrix $U$ becomes diagonal
\begin{eqnarray}
U = \mbox{diag} (-1,1,1,1,-1,-1),
\label{U-mat1}\end{eqnarray}
and our representation coincides with the one in the spinor basis.
In what follows all formulae will be derived in terms of
the partial amplitudes (\ref{spectr}) or (\ref{your}),
nevertheless the numerical calculations are performed with
our solutions (\ref{sys2}) by utilizing eqs.~(\ref{odd}) and
(\ref{U-mat}).

Coming back to the normalization condition
it is easy to show that eq.~(\ref{normdc}) may be
transformed to a diagonal form
\begin{equation}
\frac{2}{M_D}\int\!\frac{d{p_4} \, d|{\bf p}| |{\bf p}|^2}{(2\pi)^4}
(Y^+({p_4},|{\bf p}|),\hat \omega Y({p_4},|{\bf p}|))=1,\label{module}
\end{equation}
which is exactly the normalization condition used in ref.\ \cite{ZT}.
In eq.~(\ref{module}) $Y$ denotes the eight component vector (\ref{your}),
and
$\hat\omega$ is a diagonal matrix
\begin{equation}
\hat\omega=- \mbox{diag} (M_D,M_D,M_D,M_D,M_D-2E_{\bf p},2E_{\bf p}+
M_D,M_D-2E_{\bf p},2E_{\bf p}+M_D),
\label{diag}
\end{equation}
so that the integrand in eq.~(\ref{module}) consists of a sum of
quadratic terms of radial functions
$Y_\alpha$  weighted
with  $\omega_\alpha$. Therefore
each term, after integration, may be interpreted as
pseudo-probability of finding the corresponding relativistic
state in the deuteron. The result of
our numerical calculations of the pseudo-probabilities
is presented in Table~\ref{pseudover}.
It is seen that an admixture of the negative-energy
amplitudes affect the
contribution of the positive-energy states. The appearance of the negative
contributions of waves with negative
$\rho$-spin is not a surprise; it follows from
the physical meaning of the normalization condition
according to that the contribution of each term in eq.~(\ref{module})
is the effective baryon charge in  the corresponding state.
The pseudoprobabilities of $S$ and $D$ waves (see  Table~\ref{probab})
are close
to the corresponding probabilities
obtained in the   nonrelativistic
Bonn and Paris potentials, which is to be expected, since
the deuteron is essentially nonrelativistic system.

To investigate the behavior of the partial amplitudes
and their nonrelativistic
limits, we employ once more the normalization
integral (\ref{ladder-norm}),
now however in the form of eq.~(\ref{module}). Then,
similar to eqs.~(\ref{charged}) and  (\ref{spin}),
we define the following functions $\psi^\alpha $
depending upon $|{\bf p}|$ by
\begin{equation}
\psi^\alpha (|{\bf p}|)=\sqrt{2\int\!dp_4\,
\omega_\alpha |Y_\alpha (p_4,|{\bf p}|)|^2/M_D}.
\label{wv1}
\end{equation}
Thus  $\psi^\alpha$ may be regarded as the absolute value
of the relativistic wave function of the deuteron in
the state $\alpha$ (for instance, $\alpha =5$ corresponds to
$^3S_1^{++}$ configuration, $\alpha =7$  to
$^3D_1^{++}$ etc., cf.\ eq.~(\ref{your})).

Figures (\ref{sppkaz}) and (\ref{dppkaz}) display the behavior
of the relativistic wave functions $\psi_0^+$ and $\psi_2^+$
(solid lines) versus the relative momentum $|{\bf p}|$
in comparison with the nonrelativistic $S$ and $D$ waves.
We conclude that with an accuracy of model ambiguities
in the nonrelativistic calculations (cf., the difference
between Paris and Bonn wave functions, i.e.,
the dashed lines in figs.~(\ref{sppkaz}) and (\ref{dppkaz}))
the large relativistic components are  close to their
nonrelativistic analogues  up to $|{\bf p}|\sim m$.
However, there is a distinctive difference in
the shape of the $D$ waves in the two approaches.
Namely, the nonrelativistic functions   change
the sign in the region $|{\bf p}|\sim m$, whereas the BS
component does not do so (cf.  the solid line labeled as BS-I
in fig.~\ref{dppkaz}). To understand this
we tentatively introduce an auxiliary definition of the
relativistic $D$ wave which is just the difference between
the integrand in the normalization condition
and the contribution of the $^3S_1^{++}$  component, i.e.,
we introduce in the definition of $D$
the contribution of all the negative energy states:
$\tilde \psi_2^+\sim\sqrt{{w^+}^2+{w^-}^2+{u^-}^2+\ldots}$.
In this case only two wave functions $\psi_0^+$ and $\tilde \psi_2^+$
determine the normalization of the BS amplitude, and
the correspondence with the nonrelativistic limit
becomes one to one.
In fig.~\ref{dppkaz} the function $\tilde \psi_2^+$ is labeled
by BS-II, and it is seen that it displays a minimum in the
same region as the nonrelativistic functions, i.e.,
it has the same shape as the nonrelativistic $D$ wave.
One observes that the nonrelativistic
$D$ wave  already mimicks relativistic effects, so that
in calculations of relativistic corrections
to the nonrelativistic approaches an
overestimate of the magnitude of such corrections may occur.
For completeness, in fig.~\ref{pkaz} we present the
wave functions for $L=1$; since the waves $u^-$ and $w^-$ are
negligibly small, even in comparison with the waves $L=1$,
they are not presented here. 

\subsection{The vertex functions}

In studying the nonrelativistic correspondence
of the  solutions of the BS equation it is convenient to work with
the BS vertices $G(p;P)$ defined as
\begin{equation}
\chi(p;P) = \frac{(\hat p_1 + m)\, G(p;P)\,(\hat p_2 + m)}
{(p_1^2-m^2)(p_2^2-m^2)}.
\label{ver}
\end{equation}
>From eqs.~(\ref{razlozhenie}) and (\ref{ver}) it is possible to find
a partial decomposition for the vertex  $G(p;P)$. In doing so, one
introduces the two four-vectors of on-mass-shell particles corresponding
to the Dirac spinors in eq.~(\ref{spinangle}), i.e.,
\begin{equation}
k_1\,=\,(E_p,{\bf p}),\quad k_2\,=\,(E_p,-{\bf p}),\quad
E_p=\sqrt{{\bf p}^2+m^2},\quad p=(p_0,{\bf p}).
\label{freevec}
\end{equation}
Then in eq.~(\ref{ver}) the inverse propagator of the nucleons
may be represented in terms of the vectors $k_{1,2}$ by
\begin{eqnarray}
&&
S^{-1}(1)\equiv \frac{\hat P}{2}+\hat p - m=
\frac{1}{2E_p}\left[
(\hat k_1 -m) {\cal S}_-^{-1}(1)
+(\hat k_2 +m ){\cal S}_+^{-1}(1)\right ],\nonumber\\[2mm]
&&
S^{-1}(2)\equiv\frac{\hat P}{2}-\hat p -m=
\frac{1}{2E_p}\left [
(\hat k_2 -m){\cal S}_-^{-1}(2)
+(\hat k_1 +m){\cal S}_+^{-1}(2)\right ],
\label{invprop}
\end{eqnarray}
where
\begin{equation}
{\cal S}_{\pm}(1)=\left (\frac{M_D}{2}+p_0 \mp E_p\right )^{-1},\quad
{\cal S_{\pm}}(2)=\left (\frac{M_D}{2}-p_0 \mp E_p\right )^{-1}.
\label{enerprop}
\end{equation}
Because of
\begin{eqnarray}
&&
S^{-1}(1)\, U_s^{\rho_1}({\bf p})\,=\,{\rm sign} (\rho_1)\,
{\cal S}_{\rho_1}(1)\,
U_s^{\rho_1}(-{\bf p})\nonumber\\[2mm]
&&
S^{-1}(2) \,U_s^{\rho_2}(-{\bf p})\,=\,{\rm sign} (\rho_2)\,
{\cal S}_{\rho_2}(2)\,
U_s^{\rho_2}({\bf p})
\label{proj}
\end{eqnarray}
the partial decomposition of $G(p;P)$ reads
\begin{equation}
G_{\cal M}(p;P)\,=\,\sum\limits_\alpha G^\alpha (p_0,|{\bf p}|)
{\cal V}^\alpha_{\cal M} (-{\bf p}),
\label{parver}
\end{equation}
hence the partial amplitudes and the vertex functions
are interrelated via the following simple expression
\begin{equation}
Y^\alpha (p_0,|{\bf p}|) = \,{\cal S}_{\rho_1}(1)\,
{\cal S}_{\rho_2}(2)\,G^\alpha (p_0,|{\bf p}|).
\label{relatver}
\end{equation}
The relation eq.~(\ref{relatver}) implies that the BS
amplitudes  (\ref{razlozhenie}) have sharp
maxima around $p_0=0$,  while
the behavior of the partial vertices $G^\alpha (p_0,|{\bf p}|)$
is predicted to appear as smooth functions of the relative
energy (see also ref.\ \cite{ZT}).

The behavior of the vertex functions is shown
in figs.~\ref{gspp3d} and \ref{gdpp3d}
for the configurations $^3S_1^{++}$
and $^3D_1^{++}$ as functions of the relative energy
$p_0$ and
momentum $|{\bf p}|$ in the Wick-rotated system.
It is seen that the dependence of the vertex functions
upon the relative energy is weak, hence one may expect
that the nonrelativistic and relativistic vertices at $p_0=0$
have similar structures as functions of $|{\bf p}|$.
>From this observation and eq.~(\ref{relatver})
we establish another
relation between the BS amplitudes and   nonrelativistic wave
functions.  Below, as an example, we show how one can obtain
the relativistic wave function
for the $^3S_1^{++}$ configuration
from the BS amplitude. The energy dependence of the component $u^+$ is
factorized into two parts, namely a dependence on
the scalar propagators (\ref{enerprop}) and a
vertex function. Then using the smoothness of the vertex as function
of $p_0$ we replace it by its value at $p_0=0$ multiplied
by a smooth function of $p_0$, i.e.,
\begin{eqnarray}
&&
u^+ (p_0,|{\bf p}|) =
\frac{G^+(p_0,|{\bf p}|) }{\left [ (\frac{M_D}{2}-E_p)^2-p_0^2\right ] }
=
\frac{G^+(0,|{\bf p|})\,\xi(p_0,|{\bf p|}) }{
\left [ (\frac{M_D}{2}-E_p)^2-p_0^2\right ] }
\label{splus}
\end{eqnarray}
with $\xi(0,|{\bf p}|)=1$,
where the dimensionless
function $ \xi(p_0,|{\bf p}|)$ reflects the energy dependence
of the vertex function.  In view of the smooth behavior of the
vertices as function on $p_0$, one may replace this function by a constant,
$\xi(0,|{\bf p}|)\approx \xi_0$ (with $\xi_0\sim 1$).
Then in the
normalization integral eq.~(\ref{module}) the integration
over the relative energy may be carried out explicitly and
the remaining part corresponds to the
square of the nonrelativistic wave function, i.e.,
we define the nonrelativistic limit of the BS amplitude
$u^+$ as follows
\begin{equation}
\psi_0(|{\bf p}|) = \xi_0 u^+(0,|{\bf p}|)\,
\frac{(M_D/2-E_p)}{2\sqrt{M_D}}.
\label{sroed}
\end{equation}
Similar definitions, using eqs.~(\ref{module}), (\ref{diag}),
(\ref{enerprop}) and  (\ref{relatver})
are valid for other waves.
The generalized relativistic
$S$ and $D$ waves in this manner
are displayed in figs.~\ref{spgen} and \ref{dpgen}.
The actual calculations have been performed with $\xi_0=1$. A comparison
with the corresponding nonrelativistic wave functions
at $|{\bf p}|\to 0$ shows that, by choosing the  parameter
$\xi_0=1$, we slightly (by about 10\%) overestimate the
relativistic functions (see figs.~\ref{gspp3d} and \ref{gdpp3d}).
It is worth stressing that in our solution of
the BS equation the relativistic $D$ wave does not change
its sign in the interval up
to $|{\bf p}|\sim 1.5$ GeV/c. This is the most
essential difference between the relativistic and nonrelativistic
approaches in this region.
Therefore, it is expected that the relativistic
corrections to physical quantities
in the deuteron up to $|{\bf p}|\sim 1$ GeV/c,
are relatively small; to distinguish them one should either
compute observables which are known experimentally
with a very high precision and sensitive
to the spin structure,
or find special processes where
the large components are suppressed and only the states with
negative energies are relevant.

\section{The static characteristics of the deuteron}

Let us calculate  the   static
moments of the deuteron in   the  BS formalism.
The conserved electromagnetic
current of the deuteron (\ref{current}) in terms of the
BS amplitude is given by
\begin{eqnarray}
\langle P^\prime, \lambda^\prime|J_\mu|P,\lambda\rangle
=- i  e N_D \int d^4p\,
{\mbox Tr}\left\{
\bar\chi_{\lambda^\prime}(p^\prime;P^\prime)\Gamma_\mu(q)
\chi_\lambda(p;P)S_F(\hat p_2)^{-1}\right\},
\label{EMcur}
\end{eqnarray}
where $S_F(p_2)=\hat p_2+m$, $p^\prime=p+q/2$,
$P^\prime=P+q$, $N_D=1/{(2\pi)^4}/2M_D$.
The quantity $\Gamma_\mu$ is the photon-nucleon electromagnetic vertex, which
is assumed to be of the on-mass-shell form
\begin{eqnarray}
\Gamma_\mu(q)=\gamma_\mu F_1^s(q^2)-\frac{\kappa}{2m}
\sigma_{\mu\nu}q^\nu F_2^s(q^2),
\label{photvertex}
\end{eqnarray}
where $\sigma_{\mu\nu}={1\over 2}[\gamma_\mu,\gamma_\nu]$, and
$F_i^s$ are isoscalar Dirac (Pauli)
form factors of the nucleon
with $F_1(0)=F_2(0)={1/2}$,
$\kappa=\mu_p+\mu_n-1$, and $\mu_{p,n}$
are the proton and neutron anomalous magnetic
moments in units of the nuclear magneton $1/(2m)$.
The gauge invariance of the
electromagnetic current in the ladder approximation
has been proven in ref.\ \cite{ZT} (see also \cite{nagorny}).

Now we have to interrelate the expression for the static moments
(\ref{magnetic}) and (\ref{quadrupole}), which are
determined in the Breit frame, and the BS amplitudes, which are numerically
obtained in the rest frame of the deuteron.
This relation is given by the general transformation rules
\begin{eqnarray}
&&\chi_\lambda(p;P)=\Lambda({\cal L})
\chi_\lambda({\cal L}^{-1}p;P_{c.m.})
\Lambda^{-1}({\cal L}),
\label{l1}\\[2mm]
&&\bar\chi_\lambda(p^\prime;P^\prime)=
\Lambda^{-1}({\cal L})\bar\chi_\lambda({\cal L}p^\prime;
P_{c.m.})\Lambda({\cal L}),
\label{l2}\\[2mm]
&&\Lambda^{-1}({\cal L})S_F(\frac12 P-p)^{-1}\Lambda({\cal L})
=S_F(\frac12 P_{c.m.}-{\cal L}^{-1}p)^{-1},
\label{l3}
\end{eqnarray}
where
$\Lambda$ is the operator for
spin-${1\over 2}$ particles corresponding to the
Lorentz transformation $P={\cal L}P_{c.m.}$,
$P^\prime={\cal L}^{-1}P_{c.m.}$,
\begin{eqnarray}
&&\Lambda({\cal L})=\frac{M_D+\hat P\gamma_0}{\sqrt{2M_D(E+M_D)}}
\label{lorlam}
\end{eqnarray}
with the corresponding
Lorentz transformation matrix ${\cal L}$
\begin{eqnarray}
{\cal L} = \left(
\begin{array}{cccc}
\sqrt{1+\eta} & 0 & 0 & -\sqrt{\eta} \\
0 & 1 & 0 & 0 \\
0 & 0 & 1 & 0 \\
-\sqrt{\eta} & 0 & 0 & \sqrt{1+\eta}.
\end{array}
\right)
\label{mtrtransf}
\end{eqnarray}
The direction of the boost is supposed to be parallel to $q_Z$.
Then  after the Lorentz transformation of the
integrand in eq.~(\ref{EMcur}), the matrix element takes the form
\begin{eqnarray}
&&
\langle P^\prime, \lambda^\prime |J_\mu |P,\lambda\rangle  =
\nonumber \\[2mm]
&&
- i  e N_D\int d^4p\, {\mbox Tr} \left\{
\bar\chi_{\lambda^\prime}(p^\prime;P_{c.m.})
\tilde\Gamma_\mu(q^2)\chi_\lambda(p;P_{c.m.})
S_F(\frac12 P_{c.m.}-p)^{-1}
[\Lambda^{-1}({\cal L})]^2\right\},
\label{cur}
\end{eqnarray}
where
\begin{eqnarray}
\tilde\Gamma_\mu(q)=\Lambda({\cal L})\Gamma_\mu(q)\Lambda({\cal L})
\label{boostvertex}
\end{eqnarray}
and the variable $p^\prime$ is represented via $p$ and $q$ as
\begin{eqnarray}
p^{\prime} =
{\cal L} p^{\prime \ (B)} =
{\cal L} \bigl( p^{(B)} + \frac{1}{2}q) =
{\cal L}^{2} p + \frac{1}{2} {\cal L}q,
\label{ltr}
\end{eqnarray}
with components
\begin{eqnarray}
&&p^{0\prime}=(1+2\eta)p_0-2\sqrt{\eta}\sqrt{1+\eta}p^z-M_D\eta,\\
&&p^{x\prime}=p^x,\quad p^{y\prime}=p^y,\\
&&p^{z\prime}=
(1+2\eta)p^z-2\sqrt{\eta}\sqrt{1+\eta}p^0+M_D\sqrt{\eta}\sqrt{1+\eta}.
\end{eqnarray}

Eq.~(\ref{cur}) is the starting point
in evaluating the static moments of the deuteron in the BS formalism.
The main peculiarities of this matrix element, in comparison
with the familiar nonrelativistic expression,
come from the Lorentz transformation and
from the relativistic nature of the BS amplitude itself
and might be summarized by

(i) effects of the negative-energy partial states
(especially nondiagonal expectation of the current
between  $^3S_1^{++}$ and $^1P_1^{(e),(o)}$, $^3P_1^{(o),(e)}$
partial states),

(ii) a dependence of the amplitude upon the relative
energy $p_0\neq 0$; in studying the static characteristics of the deuteron
this effect is called retardation in the BS amplitude,

(iii) an effect of boosting on the internal space-time
variable, that is, the effect of ${\cal L}\neq 1$

(iv) effects of the deformation of the BS amplitude concerning the
booster $\Lambda({\cal L})\neq 1$.

In fact, in the matrix element (\ref{cur})
these boost effects reduce to a deformation of
the photon-nucleon
vertex eq.~(\ref{boostvertex}) and to corrections from
$[\Lambda^{-1}({\cal L})]^2$. In our case, i.e., as
$\eta\to 0$ (see
eqs.~(\ref{magnetic}) and (\ref{quadrupole}))
for the eqs.~(\ref{cur}) and (\ref{boostvertex}) it may be written:
\begin{eqnarray}
&&
[\Lambda^{-1}({\cal L})]^2 \simeq
1+\sqrt{\eta}\gamma_0\gamma_3+\frac{\eta}{2},
\label{squareboost}\\[2mm]
&&
\Lambda({\cal L})\gamma_0\Lambda({\cal L})=\gamma_0
\label{qqqq}\\[2mm]
&&
\Lambda({\cal L})\gamma_1\Lambda({\cal L})=\gamma_1
[\Lambda({\cal L})]^2 \label{uuuu}\\[2mm]
&&
\Lambda({\cal L})\gamma_\alpha\,\hat q\,
\Lambda({\cal L})=\gamma_\alpha\,\hat q,\quad (\alpha=0,1)
\label{wwww}
\end{eqnarray}
In what follows, the deviation of the quantity
$[\Lambda^{-1}({\cal L})]^2$ from unity in the matrix
element eq.~(\ref{cur}) we call the effects of the Lorentz
boost in the BS amplitude.

\subsection{The quadrupole moment}

\subsubsection{General formulae}

Accordingly to the eqs.~(\ref{quadrupole}),
(\ref{photvertex}) and (\ref{squareboost})-(\ref{wwww})
the result for quadrupole momentum is presented as follows
\begin{eqnarray}
Q_D&=&\sum\limits_{a,a^\prime}\sum\limits_{\rho,\rho^\prime}
\langle {a^\prime}^{\rho^\prime}|\hat Q|a^\rho\rangle\nonumber \\
 &=&
\sum\limits_{a,a^\prime}\sum\limits_{\rho,\rho^\prime}
\left[
\langle {a^\prime}^{\rho^\prime}|\hat Q_{C}|a^\rho\rangle
+\langle {a^\prime}^{\rho^\prime}|\hat Q_{C}^{LB}|a^\rho\rangle
+\langle {a^\prime}^{\rho^\prime}|\hat Q_{M}|a^\rho\rangle
+\langle
{a^\prime}^{\rho^\prime}|\hat Q_{M}^{LB}|a^\rho
\rangle\right],
\label{quadmat}
\end{eqnarray}
where the subscripts $C$ and $M$ mean the
corresponding contribution of the charge and
magnetic part of the photon-nucleon vertex
(\ref{photvertex}), and the superscript $LB$ is the contribution of the
Lorentz boost $([\Lambda({\cal L})^{-1}]^2-1)$.

The corresponding matrix elements  of the zero component of
the deuteron electromagnetic current in the definition
(\ref{quadrupole}) take the form ($\lambda =0,1)$
\begin{eqnarray}
&& J_0^{(\lambda,\lambda)}(P,P)=
\frac{e}{2M_D}\int \frac{d^4p}{ i (2\pi)^4}\,
{\mbox Tr}\left\{
\bar\chi_{\lambda}(p^\prime;P)
\gamma_0\chi_\lambda(p;P)S_F(\frac P2 -p)^{-1}\right\},
\label{QC}
\\
&& J_0^{(\lambda,\lambda)}(P,P)=
\sqrt{\eta}\frac{e}{2M_D}\int\frac{d^4p}{ i (2\pi)^4}\,
{\mbox Tr} \left\{
\bar\chi_{\lambda}(p^\prime;P)
\gamma_0\chi_\lambda(p;P)S_F(\frac P2 -p)^{-1}\gamma_0\gamma_3\right\},
\label{QCLB}
\\
&& J_0^{(\lambda,\lambda)}(P,P)=
-\frac{e}{2M_D}\frac{\kappa}{4m_N} \times
\nonumber\\
&&\hskip 2cm \int\frac{d^4p}{ i (2\pi)^4}\,
{\mbox Tr} \left\{
\bar\chi_{\lambda}(p^\prime;P)
\Lambda({\cal L}) (\gamma_0\hat{q}-\hat{q}\gamma_0) \Lambda({\cal L})
\chi_\lambda(p;P)S_F(\frac P2-p)^{-1}\right\},
\label{QM}
\\
&& J_0^{(\lambda,\lambda)}(P,P)=
-\frac{e}{2M_D}\frac{\kappa\sqrt{\eta}}{4m_N} \times
\\
&&\hskip 2cm \int\frac{d^4p}{ i (2\pi)^4}\,
{\mbox Tr} \left\{
\bar\chi_{\lambda}(p^\prime;P)
(\gamma_0\hat{q}-\hat{q}\gamma_0)
\chi_\lambda(p;P)S_F(\frac P2-p)^{-1}\gamma_0\gamma_3\right\}.
\label{QMLB}
\end{eqnarray}
As next step the partial wave decomposition of
eqs.~(\ref{QC}) - (\ref{QMLB}) has to be performed.
Then one expands the integrands in Taylor
series around $\eta=0$ and carries out the
limit $\eta \to 0$. It is clear that one has to keep
corrections including $O(\eta)$ in the wave function
$\chi(p_0^\prime,{\bf p}^\prime;P)$ and the matrix $\Lambda({\cal L})$.

This scheme of calculation allows to investigate
separately the contribution of different
relativistic effects  mentioned  above.
The eqs.~(\ref{QCLB}) and (\ref{QMLB}) are
new contributions which account for the effect of
the boosted photon-nucleon vertex. Moreover, the
Lorentz deformation effect of the BS amplitude also is
taken into account in these matrix elements through the
relative momentum $p^\prime$.

Obviously, the main contributions to the quadrupole moment comes from the
charge part
$\langle {a^\prime}^{\rho^\prime}|Q_C|a^\rho\rangle$, computed
with the large $S$ and $D$ components of the BS amplitude.
For these states,
with $\rho=\rho^\prime=+1$, one can recover the nonrelativistic
formula for the quadrupole moment of the deuteron and separate
the corrections due
to the relativistic Fermi motion of the nucleons and the retardation
in the relative energy
\begin{eqnarray}
&&Q_D^C=\sum\limits_{a,a^\prime=S,D}
\langle {a^\prime}^+|\hat Q_{C}|a^+\rangle
=Q^{(+,+)}_{p}+Q^{(+,+)}_{p_0},
\label{twoterms}
\end{eqnarray}
where the two terms in r.h.s. of the eq.~(\ref{twoterms}) reflect
the existence of derivatives in respect
to the momentum $|{\bf p}|$ and the relative energy
in the corresponding integrands:
\begin{eqnarray}
&&Q_{p}^{(+,+)}  =  -
\frac{e}{2M_D}
\int\limits_{-\infty}^{+\infty}
\int\limits_0^{+\infty}\!\frac{dp_0|{\bf p}|^2d|{\bf p}|}{ i (2\pi)^4}
(E_{\bf p}-{M_D\over 2}+p_0)\label{upwp}\\[2mm]
&&
\times\left\{
(1-{2p_0\over M_D})^2\left[-{1\over 12}
\frac{(E_{\bf p}-m)^2}{|{\bf p}|^2E_{\bf p}^2}\Upp^2\right.\right.
\nonumber\\[2mm]
&&
-{1\over 120}\frac{14E_{\bf p}^4+5E_{\bf p}^2m^2-3m^4+
20E_{\bf p}^3m}{|{\bf p}|^2E_{\bf p}^4}\Wpp^2\nonumber\\[2mm]
&&+{1\over 10}\Wpp\frac{1}{|{\bf p}|}
\frac{\partial}{\partial |{\bf p}|}{\Wpp}
+{1\over 20}\Wpp\frac{\partial^2}{\partial |{\bf p}|^2}{\Wpp}
\nonumber\\[2mm]
&&+{\sqrt{2}\over 60}
\frac{3m^4-4E_{\bf p}^4+5E_{\bf p}^2m^2+5E_{\bf p}^3m}
{|{\bf p}|^2E^4_{\bf p}}\Upp\Wpp\nonumber\\
&&+{\sqrt{2}\over 20}\frac{2E_{\bf p}+3m}{E_{\bf p}}\Upp
\frac{1}{|{\bf p}|}\frac{\partial}{\partial |{\bf p}|}{\Wpp}
\nonumber\\[2mm]
&&+{\sqrt{2}\over 20}\frac{2E_{\bf p}-3m}{E_{\bf p}}\Wpp
\frac{1}{|{\bf p}|}\frac{\partial}{\partial |{\bf p}|}{\Upp}
\nonumber \\[2mm]
&&\left.+{\sqrt{2}\over 20}\Wpp
\frac{\partial^2}{\partial |{\bf p}|^2}{\Upp}
+{\sqrt{2}\over 20}\Upp
\frac{\partial^2}{\partial|{\bf p}|^2}{\Wpp}\right]
\nonumber\\[2mm]
&&+{1\over 5}\frac{|{\bf p}|^2}{M_D^2}
\left[{3\over 2}\frac{1}{|{\bf p}|^2}\Wpp^2+\Wpp
\frac{1}{|{\bf p}|}\frac{\partial}{\partial |{\bf p}|}{\Wpp}
+\frac{3\sqrt{2}}{|{\bf p}|^2}\Upp\Wpp\right.\nonumber\\[2mm]
&&\left.\left.+\sqrt{2}\Wpp
\frac{1}{|{\bf p}|}\frac{\partial}{\partial |{\bf p}|}{\Upp}
+\sqrt{2}\Upp
\frac{1}{|{\bf p}|}
\frac{\partial}{\partial |{\bf p}|}{\Wpp}\right]\right\}.
\nonumber
\end{eqnarray}
and
\begin{eqnarray}
&&Q_{p_0}^{(+,+)}  = 
\frac{e}{2M_D}
\int\limits_{-\infty}^{+\infty}
\int\limits_0^{+\infty}\!\frac{dp_0|{\bf p}|^2d|{\bf p}|}
{ i (2\pi)^4}{1\over 5}
\frac{|{\bf p}|^2}{M_D^2}(E_{\bf p}-{M_D\over 2}+p_0)\label{Qeps}\\
&&\left\{
-\sqrt{2}
\left[\Upp\frac{\partial^2}{\partial p_0^2}{\Wpp}
+\Wpp\frac{\partial^2}{\partial p_0^2}{\Upp}\right]\right.
\nonumber\\
&&\left.-\Wpp\frac{\partial^2}{\partial p_0^2}{\Wpp}
\right\}\nonumber\\
&&+\frac{e}{2M_D}
\int\limits_{-\infty}^{+\infty}
\int\limits_0^{+\infty}\!\frac{dp_0|{\bf p}|^2d|{\bf p}|}
{ i (2\pi)^4}
{3\over 10M_D}(1-{2p_0\over M_D})^2(E_{\bf p}-{M_D\over 2}+p_0)
\nonumber\\
&&\left\{
\sqrt{2}\left[\left(1+{m\over E_{\bf p}}\right)
\Upp\frac{\partial}{\partial p_0}{\Wpp}
+\left(1+{m\over E_{\bf p}}\right)
\Wpp\frac{\partial}{\partial p_0}{\Upp}\right]\right.
\nonumber\\
&&\left.
+\Wpp\frac{\partial}{\partial p_0}{\Wpp}\right\}
\nonumber\\
&&+\frac{e}{2M_D}
\int\limits_{-\infty}^{+\infty}
\int\limits_0^{+\infty}\!\frac{dp_0|{\bf p}|^2d|{\bf p}|}
{ i (2\pi)^4}
{|{\bf p}|\over 5M_D}(1-{2p_0\over M_D})(E_{\bf p}-{M_D\over 2}+p_0)
\nonumber\\
&&\left\{
\sqrt{2}\left[\Upp\frac{\partial^2}{\partial p_0\partial |{\bf p}|}{\Wpp}
+\Wpp\frac{\partial^2}{\partial p_0\partial |{\bf p}|}{\Upp}
\right]\right.
\nonumber\\
&&\left.
+\Wpp\frac{\partial^2}{\partial p_0\partial |{\bf p}|}{\Wpp}\right \},
\nonumber
\end{eqnarray}
where $\Upp$ and $\Wpp$
represent the radial function of
the corresponding partial state $^3S_1^{++}$ and $^3D_1^{++}$.
In the nonrelativistic approximation, $E_{\bf p}\to m_N$, $p_0/M_D\to 0$,
eq.~(\ref{upwp}) yields
\begin{eqnarray}
&&Q^{(+,+)}_{p} \to \nonumber \\[2mm]
&&-\frac{e}{2M_D}\frac{1}{10\sqrt{2}}
\int\limits_{-\infty}^{+\infty}\int\limits_0^{+\infty}\!
\frac{dp_0|{\bf p}|^2d|{\bf p}|}{ i (2\pi)^4}
\left\{
\left[
\left(
-\frac{\partial^2}{\partial |{\bf p}|^2}+
\frac{1}{|{\bf p}}\frac{\partial}{\partial |{\bf p}|}
\right)\Upp\right]\Wpp\right.\nonumber\\
&&-\left.\left[\left(\frac{\partial^2}{\partial |{\bf p}|^2}+
\frac{5}{|{\bf p}}\frac{\partial}{\partial |{\bf p}|}+
\frac{3}{|{\bf p}|^2}\right)
\Wpp\right]\Upp\right\}(E_{\bf p}-\frac{M_D}{2})\nonumber\\[2mm]
&&+\frac{e}{2M_D}\frac{1}{20}
\int\limits_{-\infty}^{+\infty}\int\limits_0^{+\infty}\!
\frac{dp_0|{\bf p}|^2d|{\bf p}|}{ i (2\pi)^4}
\left[\left(
\frac{\partial^2}{\partial |{\bf p}|^2}+
\frac{2}{|{\bf p}}\frac{\partial}{\partial |{\bf p}|}
-\frac{6}{|{\bf p}|^2}\right)\Wpp\right]\nonumber \\[2mm]
&&\times \Wpp(E_{\bf p}-\frac{M_D}{2}).\label{reduc}
\end{eqnarray}

The expression eq.~(\ref{reduc}) has not yet a "true" nonrelativistic
form because of the integration over $p_0$. However, by making use of
the eqs.~(\ref{splus}), (\ref{sroed}) with $\xi_0=1$ and carrying out
the $p_0$-direction integration explicitly, the familiar
nonrelativistic  expression \cite{nonrelquad} for the quadrupole
moment is reproduced exactly
\begin{eqnarray}
Q_D & = & -\frac{1}{20}
\int\!\frac{d |{\bf p}|}{(2\pi)^3}
\left\{
\sqrt{8}
\left[|{\bf p}|^2\frac{d\psi_0(|{\bf p}|)}{d|{\bf p}|}
\frac{d\psi_2(|{\bf p}|)}{d|{\bf p}|}+3|{\bf p}|\psi_2(|{\bf p}|)
\frac{d\psi_0(|{\bf p}|)}{d|{\bf p}|}\right]\right.
\label{QNRmom}\\[2mm]
&& + \left.
|{\bf p}|^2\left(\frac{d\psi_2(|{\bf p}|)}{d|{\bf p}|}\right)^2
+6\left(\psi_2(|{\bf p}|)\right)^2\right\},
\nonumber
\end{eqnarray}
where
$\psi_0(|{\bf p}|)$ and $\psi_2(|{\bf p}|)$ are defined by
eq.~(\ref{sroed}) and correspond to the nonrelativistic
$S$ and $D$ components of the deuteron wave function
(see, also  figs.~\ref{spgen} and \ref{dpgen}).
As seen from eq.~(\ref{QNRmom}) the main contribution to the matrix
element (\ref{upwp}) is expected to come from the interference
of the positive $S$ and $D$-states in the deuteron; the remaining
terms with negative $\rho$- spins are the contribution
of the relativistic Fermi motion.

The  second term $Q_{p_0}^{(+,+)}$ in eq.~(\ref{twoterms})
and the matrix element of the Lorentz boost operator (\ref{quadmat})
are of a pure relativistic nature and reflect the relativistic
corrections to the quadrupole moment. For instance,
for the positive states the corrections
$\hat Q_{C}^{LB}$,
$Q^{(++)}_{LB}\equiv\sum\limits_{a,a^\prime=S,D}
\langle {a^\prime}^+|\hat Q_{C}^{LB}|a^+\rangle$ are
\begin{eqnarray}
Q^{(++)}_{LB} & = &
\frac{e}{2M_D}\int\limits_{-\infty}^{+\infty}
\int\limits_0^{+\infty}\!\frac{dp_0|{\bf p}|^2d|{\bf p}|}{ i (2\pi)^4}
(E_{\bf p}-\frac{M_D}{2}+p_0)(1-\frac{2p_0}{M_D})\frac{1}{5M_D}
\frac{1}{E_{\bf p}} \times \label{LB1}\\[2mm]
&& \left\{
\frac{6E_{\bf p}^2-2mE_{\bf p}-m^2}{E^2_{\bf p}}
\left[
\frac{1}{2}\Wpp^2+\sqrt{2}\Upp\Wpp\right]\right.\nonumber\\[2mm]
&&+\sqrt{2}|{\bf p}|
\left[\Upp\frac{\partial}{\partial |{\bf p}|}\Wpp
+\Wpp\frac{\partial}{\partial |{\bf p}|}\Upp\right]\nonumber\\[2mm]
&&+\left.|{\bf p}|\Wpp\frac{\partial}{\partial |{\bf p}|}
\Wpp\right\}\nonumber\\[2mm]
&&-\frac{e}{2M_D}
\int\limits_{-\infty}^{+\infty}
\int\limits_0^{+\infty}\!\frac{dp_0|{\bf p}|^2d|{\bf p}|}{ i (2\pi)^4}
(E_{\bf p}-\frac{M_D}{2}+p_0)\frac{1}{5}
\frac{|{\bf p}|^2}{M_D^2E_{\bf p}} \times \label{LB2}\\[2mm]
&& \left\{
\sqrt{2}\left[\Upp\frac{\partial}{\partial p_0}\Wpp+
\Wpp\frac{\partial}{\partial p_0}\Upp\right]\right.\nonumber\\[2mm]
&&+\left. \Wpp\frac{\partial}{\partial p_0}\Wpp\right\}.
\nonumber
\end{eqnarray}
After integration by part in Eq.~(\ref{LB1}) and (\ref{LB2}) one obtains
\begin{eqnarray}
Q^{(++)}_{LB}& = &
\frac{e}{2M_D}
\int\limits_{-\infty}^{+\infty}
\int\limits_0^{+\infty}\!\frac{dp_0|{\bf p}|^2d|{\bf p}|}{ i (2\pi)^4}
(E_{\bf p}-\frac{M_D}{2}+p_0)(1-\frac{2p_0}{M_D})\frac{2}{5M_D}
\frac{1}{E_{\bf p}} \times \label{LB3}\\[2mm]
&&\left\{
\frac{2E_{\bf p}^2-mE_{\bf p}-m^2}{E^2_{\bf p}}
\left[
\sqrt{2}\Upp\Wpp+\frac{1}{2}\Wpp^2\right]\right\}\nonumber\\[2mm]
&&
+\frac{e}{2M_D}
\int\limits_{-\infty}^{+\infty}
\int\limits_0^{+\infty}\!\frac{dp_0|{\bf p}|^2d|{\bf p}|}{ i (2\pi)^4}
\frac{1}{5}
\frac{|{\bf p}|^2}{M_D^2E_{\bf p}}
\left(1-\frac{M_D}{E_{\bf p}}\right) \times \label{LB4}\\[2mm]
&&\left\{
\sqrt{2}\Upp\Wpp+\frac12\Wpp^2\right\}.\nonumber
\end{eqnarray}
It is seen that the magnitude of this term
is of order $Q^{(++)}_{LB}\approx
\langle\frac{|{\bf p}|^2}{M_D^2}\rangle Q_{p}^{(+,+)}$ and vanishes
in the nonrelativistic limit.
In order to achieve self-consistency
in the nonrelativistic approach  to the
deuteron form factors and electrodesintegration reactions, various
relativistic corrections to the matrix elements must be
taken into account,
such as meson exchange currents and pair term
contributions \cite{mec,hocket,lock}.
In the  covariant description
of the deuteron these effects are partially accounted by
calculating transitions
between states with negative energies;
the  contribution of $P$ states in the
deuteron electromagnetic current corresponds to
diagrams with nucleon-antinucleon
pair creation in the old fashioned perturbation theory. Moreover, in
ref.\ \cite{karmanov} it has been shown that, considering the
deuteron electrodesintegration process
within the light-front dynamics, beside the dominant contribution
of expectations with  $S$ and $D$ waves,
an extra matrix element with transitions between
positive and negative energy states
is relevant to describe the electrodesintegration amplitude.
It has been also shown that the contribution of this
extra component exactly reproduces the pair term corrections
in the nonrelativistic limit.
An investigation of the correspondence between the light-front dynamics
approach  and the BS amplitude has shown \cite{dorkin}
that the extra component in ref.\ \cite{karmanov} may be imitated by
transitions between a linear combination of the $P$-waves
and $S$ or $D$ waves. Hence
in our calculation the pair terms are taken into account via
calculations of off diagonal expectation values of the relevant current
between the $S$ and $P$ partial states
(see also discussions in
refs.\ \cite{honzawa,lommon}). A more detail
analysis of the nonrelativistic limit of the expression for the
quadrupole moment with keeping leading
corrections $\sim 1/m$ will presented elsewhere.

\subsubsection{Numerical results}

The total expression for the quadrupole
moment   consists of
a multitude of terms likewise eq.~(\ref{upwp}) with
quadratic combinations of partial states and terms with
second derivatives  $\partial^2 /\partial|{\bf p}|^2$,
$\partial^2/\partial p_0^2$ and mixed ones
$ \partial^2 /\partial|{\bf p}|\partial p_0 $
computed between different partial BS amplitudes.
Their analytical form has been evaluated by an
algebraic formula manipulation code.
Concrete numerical calculations have been performed by using our
solutions of the BS equation for the partial amplitudes
eq.~(\ref{sys2}) and the relations (\ref{odd}), (\ref{U-mat}).
We find that the main contribution to the deuteron quadrupole
moment gives the first term in eq.~(\ref{quadmat}) and that
the transitions between energy even-states dominate, i.e.,
\begin{eqnarray}
Q_p^{++}=0.2690 \, {\rm fm^2}.
\label{num1}
\end{eqnarray}
This  contribution is below the experimental data
$Q_D=(0.2859 \pm 0.0003)\, {\rm fm^2}$ \cite{qexp}
by about 6\%, nevertheless it is larger than
the usual nonrelativistic calculations. This is an
understandable effect because of the specific feature of the solution
of the BS equation for which  the sum of the
pseudo-probabilities of the positive $S$ and
$D$ waves is larger than 1. In this context, since the
pseudoprobabilities of the remaining configurations are negative,
the  transitions with $P$ waves are expected to play an important role
in studying the static characteristics of the deuteron.
A particular interests present calculations of off-diagonal
expectation
between the $S$ and $P$ partial states, which is predicted
to replace the meson exchange contribution
in nonrelativistic calculations \cite{karmanov,dorkin}.
Indeed, our numerical result points to a significant
contribution of the mentioned matrix elements in
comparison with other nondiagonal transitions
\begin{equation}
\langle u^+|\hat Q_C|v_s^e\rangle = 0.0052\, {\rm fm^2};
\quad\quad
\langle u^+|\hat Q_C|v_t^o\rangle = -0.0027\, {\rm fm^2}
\label{num2}
\end{equation}
(for example, among other nondiagonal matrix elements the
largest one is
$\langle w^+|\hat Q_C|v_s^e\rangle = -0.00007 \,{\rm fm^2}$).

The part of the quadrupole moment with odd (diagonal and nondiagonal
expectations) gives a small negative contribution to the
first term in eq.~(\ref{quadmat}):
$\langle |\hat Q_C|\rangle_{odd} = -0.0007 \, {\rm fm^2}$
Gathering together all the contributions we obtain
$Q_C = 0.2706 \, {\rm fm^2}$.

An estimate  of the corrections owing to the dependence on the
relative energy,  eq.~(\ref{twoterms}), shows that they
are rather small: $Q_{p_0}^{(++)}=0.0006\, {\rm fm^2}$.

The Lorentz boost corrections have been calculated
and they are found to be negative.
Their total contribution is $Q^{LB}_C = -0.0029 \, {\rm fm^2}$ which,
together with $Q_C$, gives the final result for the electric part of the
quadrupole moment of the deuteron $Q =0.2683$.
An important moment should be stressed here. The contribution
of the Lorentz boost terms with nondiagonal transitions
between $S$ and $P$ waves are of the same order of magnitude
as those in eq.~(\ref{num2}) but opposite in sign
\begin{equation}
\langle v^+|\hat Q_C^{LB}|v_s^e\rangle = -0.0053 \, {\rm fm^2};
\quad\quad
\langle v^+|\hat Q_C^{LB}|v_t^o\rangle = 0.0027 \, {\rm fm^2}.
\label{num3}
\end{equation}
Equations (\ref{num2}) and (\ref{num3}) show that the
contribution  of  pair creation terms in nonrelativistic
calculations is predicted to
be negligibly small for the quadrupole
moment  and confirm the qualitative results obtained
in ref.\ \cite{honzawa}.

Note that our classification of the matrix elements
into the main part and Lorentz boost corrections
(cf.\ eq.~(\ref{quadmat}))
is rather conventional and does not reflect directly the
contribution of relativistic effects. However, by using
eqs.~(\ref{twoterms}), (\ref{reduc}) and (\ref{QNRmom}) we may
present our results in the form
\begin{equation}
Q_D = Q_{NR} + \delta Q_{rel.} = (0.2690 - 0.0007) \, {\rm fm^2},
\label{num4}
\end{equation}
where $Q_{NR}$ is determined by the large components of
the BS amplitude and does not depend upon the derivatives
in respect to the relative energy and upon the Lorentz boost
effects;
$\delta Q_{rel.}$ is the contribution of
all the remaining terms and, obviously,
is of a pure relativistic nature.
It is seen that the relativistic corrections to the
quadrupole moment are negative and reinforce the discrepancy,
although their magnitude is rather small. A similar conclusion
has been drawn in ref.\ \cite{honzawa} from a qualitative
analysis of the  deuteron moment within the BS formalism.

Another source of the relativistic corrections is the contribution
of the magnetic part of the effective current (\ref{quadmat}) which
vanishes in the nonrelativistic limit. Our calculation show
that its contribution to the quadrupole moment is negative too,
$\langle\, |\hat Q_M|\,\rangle = -0.0005 \, {\rm fm^2}$, so that
our final result for the deuteron quadrupole moment is
$Q_D= 0.2678\,{\rm fm^2}$, i.e.,
the discrepancy in $Q_D$ is about 6\%.

\subsection{The magnetic moment}

\subsubsection{General formulae}

Accordingly to the eqs.~(\ref{magnetic}), (\ref{EMcur})
and (\ref{squareboost})-(\ref{wwww}),
the result for magnetic moment can be written as follows
\begin{eqnarray}
&&\mu_D=\mu_{+} + \mu_{1-} + \mu_{2-} + \mu_{3-},
\label{mmain}
\end{eqnarray}
where the matrix elements between states with
positive  energies in eq.~(\ref{mmain}) are labeled by the
subscript $+$, and the subscript $-$ means that the corresponding
matrix element implements at least one wave with negative energy.
These matrix elements $\mu_{i-}$
reflect the relativistic corrections.
In order to emphasize
the nonrelativistic analogue of the magnetic moment
in the expression for the $\mu_{+}$ we subtract the
corresponding nonrelativistic
formula and the remaining part we denote as $R_+$, the
relativistic corrections due to the Fermi motion effects.
Then the functions $\mu_{\cdots}$ can be represented by
\begin{eqnarray}
&&\mu_+=(\mu_p+\mu_n)(P_{u^+}+P_{w^+}) -
\frac{3}{2}(\mu_p+\mu_n-\frac{1}{2})P_{w^+} + R_{+},
\label{muplus}\\[2mm]
&&\mu_{1-}=\frac{1}{2}(\mu_p+\mu_n)(P_{v_t^e}+P_{v_t^o})+
\frac{1}{4}(P_{v_t^e}+P_{v_t^o}) + \frac{1}{2}(P_{v_s^e}+
P_{v_s^o}) + R_{1-},
\label{mu1minus}\\[2mm]
&&\mu_{2-}=-(\mu_p+\mu_n)P_{u^-}+P_{u^-}+
\frac{1}{2}(\mu_p+\mu_n)P_{w^-} - \frac{5}{4}P_{w^-}+
R_{2-},
\label{mu2minus}\\[2mm]
&&\mu_{3-}= \sum_{a,b} C^{a,b},
\label{mu3minus}
\end{eqnarray}
where $a=u^+,w^+,u^-,w^-$, $b=v_s^e,v_s^o,v_t^e,v_t^o$, and
$P_i$ are the pseudo-probabilities of the corresponding partial
state. In eqs.~(\ref{muplus})-(\ref{mu2minus}) the diagonal
expectations between states with $L=0,2,1$ are written explicitly;
the off-diagonal contributions are included into the terms
 $R$  and $\mu_{3-}$, where
\begin{eqnarray}
R_{+} &=& -\frac{1}{3}(\mu_p+\mu_n-1+\frac{2m}{M}) H_1^{u^+} -
\frac{m}{M} H_2^{u^+} - \frac{m}{M} H_3^{u^+} - (1-\frac{2m}{M}) P_{u^+}
\nonumber\\[2mm]
&&-\frac{1}{6}(\mu_p+\mu_n-1-\frac{4m}{M}) H_1^{w^+} -
\frac{m}{M} H_2^{w^+} - \frac{m}{M} H_3^{w^+} -
\frac{1}{4} (1-\frac{2m}{M}) P_{w^+}
\nonumber\\[2mm]
&&+\frac{\sqrt{2}}{3} (\mu_p+\mu_n-1-\frac{m}{M}) H_1^{u^+,w^+},
\label{Rplus}\\[2mm]
R_{1-} &=& - \frac{1}{2}(1-\frac{2m}{M}) (\mu_p+\mu_n+\frac{1}{2})
(P_{v_t^e} + P_{v_t^o}) -
\frac{1}{2}(1-\frac{2m}{M}) (P_{\Pe} + P_{\Po})
\nonumber\\[2mm]
&& +\frac{2}{5}(2H_4^{v_t^o,v_t^e}-H_8^{v_t^o,v_t^e})
+\frac{1}{5}H_9^{v_t^o,v_t^e}
-\frac{2}{5}H_{10}^{v_t^o,v_t^e}
\nonumber\\[2mm]
&& +\frac{2}{5}(H_4^{v_s^e,v_s^o}+2H_8^{v_s^e,v_s^o})
-\frac{2}{5}H_9^{v_s^e,v_s^o}
+\frac{4}{5}H_{10}^{v_s^e,v_s^o}
\nonumber\\[2mm]
&& + \sqrt{2} (\mu_p+\mu_n-1+\frac{4m^2}{M^2})H_5^{v_t^e,v_s^o}
\nonumber\\[2mm]
&& + \sqrt{2} (\mu_p+\mu_n-1)H_5^{v_t^o,v_s^e} - \frac{\sqrt{2}}{2}
H_6^{v_t^o,v_s^e} - 2\sqrt{2} H_7^{v_t^o,v_s^e},
\label{R1minus}\\[2mm]
R_{2-} &=& - \frac{1}{3}(\mu_p+\mu_n-1-\frac{2m}{M}) H_1^{u^-} -
\frac{m}{M} H_2^{u^-} + \frac{m}{M} H_3^{u^-}
\nonumber\\[2mm]
&& - \frac{1}{6}(\mu_p+\mu_n-1+\frac{4m}{M}) H_1^{w^-} -
\frac{m}{M} H_2^{w^-} + \frac{m}{M} H_3^{w^-}
\nonumber\\[2mm]
&& + \frac{3}{4} (1-\frac{2m}{M}) P_{w^-} +
\frac{\sqrt{2}}{3} (\mu_p+\mu_n-1+\frac{m}{M}) H_1^{u^-,w^-}.
\label{R2minus}
\end{eqnarray}
The quantities $C^{a,b}$ and $H_i^{\alpha^\prime,\alpha}$
are given in the Appendix II.
Now the nonrelativistic formula for the magnetic moment
may be recovered exactly by rewriting the
term $\mu_+$ in the form
\begin{eqnarray}
&& \mu_+=\mu_{NR} + \Delta \mu_+,
\label{mmainplus}
\end{eqnarray}
where
\begin{eqnarray}
\mu_{NR} = (\mu_p+\mu_n)-\frac{3}{2}(\mu_p+\mu_n-\frac{1}{2})P_{D}
\nonumber\end{eqnarray}
reproduces the  nonrelativistic formula and the relativistic
corrections due to the Fermi motion effects are
\begin{eqnarray}
\Delta \mu_+=R_+ - (\mu_p+\mu_n)
(P_{u^-}+P_{w^-}+P_{v_s^e}+P_{v_s^o}+P_{v_t^e}+P_{v_t^o}).
\label{mnorm}
\end{eqnarray}

Finally,  the total contributions
to the  deuteron magnetic moment read as
\begin{eqnarray}
&& \mu_D=\mu_{NR} + \Delta \mu,
\nonumber\\[2mm]
&& \Delta \mu=R_{+} + \Delta \mu_{-} + \mu_{3-},
\label{mmmain}\\[2mm]
&& \Delta \mu_{-} = -(\mu_p+\mu_n)\Bigl[\frac{1}{2}(P_{v_t^e}+P_{v_t^o})+
(P_{v_s^e}+P_{v_s^o})+2P_{u^-}+\frac{1}{2}P_{w^-}\Bigr]
\nonumber\\[2mm]
&& \hskip 13.5mm +\frac{1}{4}(P_{^3P_1^{e}}+P_{v_t^o})
+ \frac{1}{2}(P_{^1P_1^{e}}+P_{^1P_1^{o}})
+ P_{u^-} - \frac{5}{4}P_{w^-} + R_{1-} + R_{2-}.
\label{deltamom}
\end{eqnarray}

\subsubsection{Numerical results}

Explicit numerical calculations give
for  the total deuteron magnetic moment
the value $\mu_D=0.856140\ (e/2m)$ which differs
from the experimentally known moment
$\mu_{exp}=0.857406\pm 10^{-6}\ (e/2m)$ \cite{honzawa}
by less then $0.15\%$.
This result consists of the nonrelativistic contribution plus
relativistic corrections listed below

(i) the main correction to the nonrelativistic value of the magnetic moment
$\mu_{NR}=0.850718\ (e/2m)$ that
comes from the transitions between positive
energy states and $P$-states ($v_s^e,v_s^o,v_t^e,v_t^o$) (the
term $\mu_{3-}$ in eq.~(\ref{mmmain}));
it gives $\mu_{3-}=6.099\cdot 10^{-3}$ and contains
$\sim 0.71\%$ of the total magnetic moment;
(ii) relativistic corrections from the
expectation values of positive energy states
of the Lorentz transformation of the intrinsic variables in the
BS amplitude (the term $R_+$ in eq.~(\ref{mmmain})), which
is found to be negative, i.e.,
$R_+=-9.75\cdot 10^{-4}$;
(iii)
the term $\Delta\mu_{-}$, and  the sum of transitions between states with
negative energy ($u^-,w^-$), and 
transitions between $P$-states themselves,
and part coming from normalization effects (eq.~(\ref{mnorm}));
this is a positive contribution with
$\Delta \mu_{-}=2.99\cdot 10^{-4}$ to the total moment.

An analysis of our numerical results
obtained for the off-diagonal expectation values
between the $S$ and $P$ partial states shows that in
contrary with eqs.~(\ref{num2})-(\ref{num3}), the contribution
of terms like pair-creation corrections in this case
do not compensate each other and give a total contribution
to the magnetic moment $\sim 0.35\%$, which is almost
$50\%$ of the total relativistic correction.

\section{Concluding remarks}

In this paper we have investigated in some detail
the numerical solution  of the Bethe-Salpeter
equation \cite{umnikov-khanna}
with a realistic one-boson-exchange interaction.
Special attention has been paid to a study of the relation
of the partial BS amplitudes to the nonrelativistic wave functions
and   to the covariant description of the static characteristics
of the deuteron.
In our analysis we consider various basises
used in defining the partial BS amplitudes
and the transition  from one basis to another.
The representation based on the complete
set of the Dirac $\gamma$-matrices and their
bilinear combinations  is found to be extremely
convenient in computing
the deuteron observables and
processes with the deuteron \cite{praha} since in this case the
dependence on the kinematical variables is mostly included in
the definition of the partial amplitudes (except for one spinor
propagator, which  usually appears when computing diagrams for
concrete processes, see ref.\ \cite{umnikov-khanna})
and the matrix structure
of the corresponding matrix element is almost independent on the
intrinsic deuteron variables. However, in this representation
an analysis of the deuteron structure in terms of
familiar $S$, $D$, etc. components and an investigation
of the  correspondence of the obtained results
with their nonrelativistic analogues is straitened. For this
sake it is more convenient to use the $\rho$ spin classification of
the amplitudes for which a physical treatment of results is
easier. In order to combine the advantages of these two representations
the corresponding unitary transformation has been presented explicitly,
cf.\ eq.~(\ref{U-mat}).  With this a hand, calculations of various processes
can be performed easily in the basis of the  Dirac matrixes and
the final expression may be treated in terms of the
$\rho$-spin partial amplitudes by utilizing eq.~(\ref{U-mat}).
This scheme of calculation has been employed in order to
compute the pseudoprobabilities of different partial states and
to find the nonrelativistic limit of the amplitudes.
In Section III different methods of comparison of our amplitudes with the
nonrelativistic $S$ and $D$ waves are presented. Apparently,
the most appropriate way to define the nonrelativistic limits
of the BS amplitudes is to use the relation (\ref{sroed}), which
is based on an analysis of the behavior of the BS vertex functions
in dependence on $p_0$ and $|{\bf p}|$ and on the nonrelativistic
relation between the vertices and wave functions in the momentum
space.  Numerical results, displayed on figs. \ref{spgen} and \ref{dpgen},
show that the generalized BS wave functions (\ref{sroed}) are
close to the nonrelativistic ones only for moderate values of $|{\bf p}|$,
while a difference occurs for $|{\bf p}| \ge m$. This means that
for rough estimates of possible relativistic effects one may
calculate the corresponding nonrelativistic expressions by
utilizing the wave functions (\ref{sroed}) instead of the
nonrelativistic $S$ and $D$ waves. Obviously, for a consistent
investigation of the relativistic corrections it is necessary
to use the covariant calculations with complete  BS amplitudes.

We have investigated the quadrupole and magnetic moments of the
deuteron within the BS formalism by computing in the Breit frame
the matrix elements of the electromagnetic current of the deuteron.
In our analysis we considered all the possible relativistic effects
connected with the Lorentz transformation from the rest frame
of the deuteron to the
Breit frame and with the dependence of the amplitude on the relative
energy $p_0$.  By utilizing results of the investigation of the
properties of the BS amplitudes performed in the Section III
and their nonrelativistic limits,
the static moments of the deuteron have been presented as a sum
of two terms, one of them having a direct nonrelativistic
analogue, the other one being  of a pure relativistic nature.
We pay special attention to the contribution of the
nondiagonal expectation values between $S$ and $P$  configurations
which are thought to include into the relativistic
calculations the  effects of pair currents widely discussed in
nonrelativistic theories. It has been shown that
for the quadrupole moment the different partial transitions
between $S$ and $P$ components possess
a noticeable magnitude, however, their summed contribution
is found to be negligibly small 
(see eqs.~(\ref{num2}) and (\ref{num3})),  whereas for the magnetic moment
these matrix elements give almost 50\% of the relativistic
effects. We obtain a good description of the experimental
data for the magnetic moment. As for the quadrupole moment
the computed value is below the experimental data by
about 6\%. That indicates that even a consistent relativistic
computation does not perfectly describe the data in the
impulse approximation. Probably, an adjustment of the operator
of the electromagnetic
current of the deuteron is needed, e.g., by including additional terms
not accounted for within the present approach such as
meson exchange currents with two-meson exchange diagrams or
$\Delta$ isobars \cite{lommon}.

\section{Summary}

In summary an analysis of the properties of the partial
Bethe-Salpeter amplitudes, obtained as numerical
solution of the BS equation with a realistic interaction,
has been performed. In order to compare relativistic amplitudes
with the nonrelativistic wave functions a method, based
on the comparative analysis of the observables, has been developed.
The static characteristics of the
deuteron, i.e., the quadrupole and magnetic moments,
have been computed within the Bethe-Salpeter formalism
with satisfactory accuracy.

\section{Acknowledgments}

We would like to thank S.M. Dorkin, F. Gross, and J. Tjon
for enlightening discussions.
L.P.K. thanks for the warm hospitality of nuclear theory group
in Research Center Rossendorf.
This research is supported in part by the National Sciences and Engineering
Research Council of Canada and by the BMBF grant 06DR666.
A.Yu.U., K.Yu.K, and L.P.K.
express their thanks to the Theoretical
Physics Institute, Univ. of Alberta, Canada
where a bulk of this work was done.

\section*{Appendix I}

The matrix form of
the spin-angular functions ${\cal V}^\alpha_{\cal M}({\bf p})$,
eq.~(\ref{spinangle}), may be obtained explicitly by replacing
the outer product of the free nucleon spinors
$U_{s_i}^{\rho}({\bf p})$ by their direct product,
$U_{s_1}^{\rho_1}({\bf p}) \otimes  U_{s_2}^{\rho_2 T}(-{\bf p})$.
The BS amplitude takes then the form
\begin{equation}
\chi_D(p_0, {\bf p}) U_C\, = \sum\limits_\alpha
\phi_\alpha (p_0, |{\bf p}|) {\Gamma}_{\cal M}^{\alpha}({\bf p}) U_C,
\nonumber
\end{equation}
with ${\Gamma}_{\cal M}^{\alpha}({\bf p})$
\begin{equation}
{\Gamma}^\alpha_{\cal M}({\bf p})=i^L
\sum\limits_{s_1 s_2 m}(LmSs|J{\cal M})(\frac{1}{2}s_1
\frac{1}{2}s_2|Ss) Y_{Lm}(\hat {\bf p}) U_{s_1}^{\rho_1}({\bf p})
U_{s_2}^{\rho_2\, T}(-{\bf p}),
\nonumber\end{equation}
where $U_C$ is the charge conjugation matrix, $U_C=i\gamma_2\gamma_0$.

One can  exploit the $\rho$ spin dependence and replace
${\Gamma}^\alpha_{\cal M}({\bf p}) \equiv
{\Gamma}^{\tilde\alpha,\,\rho_1\rho_2}_M({\bf p})$, where
\begin{eqnarray}
{\Gamma}^{\tilde\alpha,\, ++}_{\cal M}({\bf p}) &=
&\frac{\hat k_1 + m}{\sqrt{2{{E_{\bf p}}}(m+{{E_{\bf p}}})}}\;
\frac{1+\gamma_0}{2}\;
{\tilde \Gamma}^{\tilde\alpha}_{\cal M}({\bf p},\mbf{\xi})\;
\frac{\hat k_2 - m}{\sqrt{2{{E_{\bf p}}}(m+{{E_{\bf p}}})}},
\nonumber\\
{\Gamma}^{\tilde\alpha,\, --}_{\cal M}({\bf p}) &=
&\frac{\hat k_2 - m}{\sqrt{2{{E_{\bf p}}}(m+{{E_{\bf p}}})}}\;
\frac{-1+\gamma_0}{2}\;
{\tilde \Gamma}^{\tilde\alpha}_{\cal M}({\bf p},\mbf{\xi})\;
\frac{\hat k_1 + m}{\sqrt{2{{E_{\bf p}}}(m+{{E_{\bf p}}})}},
\label{gf1}\\
{\Gamma}^{\tilde\alpha,\, +-}_{\cal M}({\bf p}) &=
&\frac{\hat k_1 + m}{\sqrt{2{{E_{\bf p}}}(m+{{E_{\bf p}}})}}\;
\frac{1+\gamma_0}{2}\;
{\tilde \Gamma}^{\tilde\alpha}_{\cal M}({\bf p},\mbf{\xi})\;
\frac{\hat k_1 + m}{\sqrt{2{{E_{\bf p}}}(m+{{E_{\bf p}}})}},
\nonumber\\
{\Gamma}^{\tilde\alpha,\, -+}_{\cal M}({\bf p}) &=
&\frac{\hat k_2 - m}{\sqrt{2{{E_{\bf p}}}(m+{{E_{\bf p}}})}}\;
\frac{1-\gamma_0}{2}\;
{\tilde \Gamma}^{\tilde\alpha}_{\cal M}({\bf p},\mbf{\xi})\;
\frac{\hat k_2 - m}{\sqrt{2{{E_{\bf p}}}(m+{{E_{\bf p}}})}},
\nonumber
\end{eqnarray}
with $\tilde\alpha\in\{L,S,J\}$.

The spin-angular structures for concrete partial waves are shown in
Table~\ref{tab:3s1}.
Here $ \mbf{\xi}_{\cal M}$ is the polarization vector of the deuteron with
the components in the rest frame given by
\begin{eqnarray}
\mbf{\xi}_{+1}=(-1,-i,0)/\sqrt{2}, \quad
\mbf{\xi}_{-1}=(1,-i,0)/\sqrt{2}, \quad
\mbf{\xi}_{0}=(0,0,1).
\label{vecpol}
\end{eqnarray}
and the four-vector $\xi_{\cal M} = (0,\mbf{\xi}_{\cal M})$.

\section*{Appendix II}

Below we list the explicit form of
the quantities $C^{a,b}$ and
$H_i^{\alpha^\prime,\alpha}$ in eq.~(\ref{mmain}).
By introducing new functions $G_i^{a,b}$ the mentioned quantities
are expressed as follows:
\begin{eqnarray}
&& \tau=u^{\pm},v_s^e: \qquad C^{\tau} =
\frac{\sqrt{6}}{12}\frac{m}{M}
\Biggl[
G_1^{\tau} + 4G_2^{\tau} - 4G_3^{\tau} - G_4^{\tau}-4G_5^{\tau}\Biggr],
\nonumber\\
&& \tau=u^{\pm},v_s^o: \qquad C^{\tau} =
\frac{\sqrt{6}}{15}\frac{m}{M}
\Biggl[
- G_6^{\tau} \mp G_7^{\tau} \mp G_8^{\tau} \pm G_9^{\tau} 
- G_{27}^{\tau}\Biggr],
\nonumber\\
&& \tau=w^{\pm},v_s^e: \qquad C^{\tau} =
\frac{\sqrt{3}}{12}\frac{m}{M}
\Biggl[G_{10}^{\tau} + G_{11}^{\tau} + 4G_{3}^{\tau} + G_{4}^{\tau} +
4G_{5}^{\tau}\Biggr],
\nonumber\\
&& \tau=w^{\pm},v_s^o: \qquad C^{\tau} =
\frac{\sqrt{3}}{15}\frac{m}{M}
\Biggl[G_{12}^{\tau} \pm G_{13}^{\tau} \pm G_{14}^{\tau} \mp G_{15}^{\tau}
+ G_{28}^{\tau} \Biggr],
\nonumber\\
&& \tau=u^{\pm},v_t^e: \qquad C^{\tau} =
\frac{\sqrt{3}}{15}\frac{m}{M}
\Biggl[
\pm G_{16}^{\tau} - G_{17}^{\tau} + G_8^{\tau} - G_{9}^{\tau}
\pm G_{27}^{\tau} \Biggr],
\nonumber\\
&& \tau=u^{\pm},v_t^o: \qquad C^{\tau} =
\mp \frac{\sqrt{3}}{3}\frac{m}{M}
\Biggl[
G_{20}^{\tau} + G_{21}^{\tau} + G_3^{\tau} + \frac14 G_{4}^{\tau} +
G_{5}^{\tau}
\Biggr]
\mp \frac{\sqrt{3}}{3}\kappa G_{22}^{\tau},
\nonumber\\
&& \tau=w^{\pm},v_t^e: \qquad C^{\tau} =
\frac{\sqrt{6}}{15}\frac{m}{M}
\Biggl[
\pm G_{23}^{\tau} - G_{24}^{\tau} + G_{18}^{\tau} - G_{19}^{\tau}
\pm G_{29}^{\tau} \Biggr],
\nonumber\\
&& \tau=w^{\pm},v_t^o: \qquad C^{\tau} =
\mp \frac{\sqrt{6}}{3}\frac{m}{M}
\Biggl[
G_{25}^{\tau} + G_{26}^{\tau} + G_3^{\tau} + \frac14 G_{4}^{\tau} +
G_{5}^{\tau}
\Biggr] \pm \frac{\sqrt{6}}{6} \kappa G_{22}^{\tau},
\nonumber
\end{eqnarray}
where the $G_{i}^{\alpha,\alpha^\prime}$ are integrals of the
form
\begin{eqnarray*}
N \int dp_4 \nrmp^2 d\nrmp A_{i}(p_4,\nrmp)
[ B_{i} Y_{\alpha}(p_4,\nrmp)] Y_{\alpha^{\prime}}(p_4,\nrmp),
\end{eqnarray*}
and $A_i (p_4,\nrmp)$ are scalar functions;
$B_i $ may be either a differential operator
of the type $\partial/\partial p_4$, $\partial/\partial \nrmp$
or a scalar function ($p_4 = - i p_0$), which are summarized
in the following tabular form

\[
\begin{array}{|c|c|c|}
\hline\hline
G_i & A_{i}(p_4,\nrmp) &  B_{i} \\
\hline
1  & (E-m)Mm/(\nrmp E^2) & -\omega_{\alpha} \\
2  & -(E-m)mp_4^2/(\nrmp E^2) & 1 \\
3  & \nrmp p_4 & \partial / \partial p_4 \\
4  & M & -\omega_{\alpha} \partial/\partial p_4 \\
5  & -p_4^2 & \partial/\partial \nrmp \\
6  & (E-m)(2E+3m)p_4 /(\nrmp E) & 1 \\
7  & (E-m)(E^2+2mE+2m^2)Mp_4 /(\nrmp E^3) & 1 \\
8  & \nrmp(3E+2m)/(2E) & -\omega_{\alpha} \partial/\partial p_4 \\
9  & (3E+2m)Mp_4/E & \partial/\partial \nrmp \\
10 & (2E+m)mM/(\nrmp E^2) & -\omega_{\alpha} \\
11 & -(2E+m)m p_4^2/(\nrmp E^2) & 1 \\
12 & (2E^2-2mE-3m^2)p_4/(\nrmp E) & 1 \\
13 & (E^3-2mE^2+4m^3)Mp_4/(\nrmp E^3) & 1 \\
14 & (3E-4m)\nrmp/(2E) & -\omega_{\alpha} \partial/\partial p_4 \\
15 & (3E-4m)Mp_4/E & 1 \\
16 & (E-m)(7E+3m)p_4/(\nrmp E) & 1 \\
17 & 2(E-m)M(2E^2-mE-m^2)p_4/(\nrmp E^3) & 1 \\
18 & \nrmp(3E-m)/(2E) & -\omega_{\alpha} \partial/\partial p_4 \\
19 & (3E-m)p_4M/E & \partial/\partial \nrmp \\
20 & (E-m)^2M/(4\nrmp E^2) &  -\omega_{\alpha} \\
21 & -(E-m)^2 p_4^2 /(\nrmp E^2) & 1 \\
22 & \nrmp/(2E) & -\omega_{\alpha} \\
23 & (7E^2+2mE-3m^2)p_4 /(\nrmp E) & 1 \\
24 & (4E^3+3mE^2-m^3)Mp_4/(\nrmp E^3) & 1 \\
25 & (E^2+mE+m^2)M/(4\nrmp E^2) & -\omega_{\alpha} \\
26 & -(E^2+mE+m^2)p_4^2/(\nrmp E^2) & 1 \\
27 & (3E+2m)p_4 & \partial/\partial\nrmp \\
28 & (3E-4m)p_4 & \partial/\partial\nrmp \\
29 & (3E-m)p_4 & \partial/\partial\nrmp \\
\hline\hline
\end{array}
\]

Analogously, the functions
 $H_{i}^{\alpha,\alpha^{\prime}} (H_{i}^{\alpha,\alpha} \equiv
H_{i}^{\alpha})$ are of the same structure as
$G_{i}^{\alpha,\alpha^{\prime}}$ with \\
\[
\begin{array}{ccc||ccc}
\hline\hline
i & A_{i}(p_4,\nrmp) &  B_{i} & i & \quad A_{i}(p_4,\nrmp) & B_{i} \\
\hline
1 & \frac{1}{2}(1-m/E) & \omega_{\alpha} &
6 & M m^2/E^2 & 1 \\
2 & \frac{1}{2}(1-M/2E) & \omega_{\alpha} &
7 & -p_4^2 m/E^2 & 1 \\
3 & -p_4/E & 1 &
8 & p_4 m^3/E^3 & 1 \\
4 & p_4 m/E & 1 &
9 & \nrmp^2 m/E & \partial/\partial p_4 \\
5 & 1/2 & \omega_{\alpha} &
10 & \nrmp p_4 m/E & \partial/\partial \nrmp\\
\hline\hline
\end{array}
\]
 
\newpage
\mediumtext

\begin{table}[h]
\caption{{\em The deuteron partial amplitudes and their
transformation properties.}}
\protect\label{ampBS}
\begin{center}
\begin{tabular}{|c|c|c|c|c|c|c|c|c|} \hline
$^{2S+1}L_1$&
${\sf P}_1$ & ${\sf A}_1^0$ & ${\sf V}_1$ & ${\sf A}_0$ & ${\sf A}_2$ &
${\sf T}_1^0$ & ${\sf T}_0$ & ${\sf T}_2$ \\ \hline
$L$&1&1&1&0&2&1&0&2\\ \hline
$S$&0&0&1&1&1&1&1&1 \\ \hline\hline
${\cal K}$&
$+$&$+$&+&$+$&$+$&$-$&$+$&$+$ \\ \hline
$\Pi$ & $+$ & $-$ &+ &$+$&$+$&$-$&$+$&$+$ \\ \hline
\end{tabular}
\end{center}
\end{table}


\phantom{.}      \vskip 1cm

\begin{table}[h]
\caption{{\em The pseudo-probabilities of the partial waves in the deuteron}}
\protect\label{pseudover}
\begin{center}
\begin{tabular}{|c|c|c|c|c|}
\hline
\label{probab}
wave  & $u^{+}$ & $w^{+}$ & $u^{-}$ & $w^{-}$  \\ \hline
$P_\alpha(\%)$ & $95.014$ & $5.106$ & $-0.002$ &$-0.003$ \\ \hline \hline
wave & $v_s^e$ & $v_t^o$ & $v_s^o$ & $v_t^e$ \\ \hline
$P_\alpha(\%)$ & $-0.010$ & $-0.082$ & $-0.015$ & $-0.008$ \\ \hline
\end{tabular}
\end{center}
\end{table}


\begin{table}[h]
\caption{{\em
Spin-angular functions $\tilde \Gamma_{\cal M}^{\tilde\alpha}$
for the deuteron channel.}}
\protect\label{tab:3s1}
\[
\begin{array}{cc}
\hline\hline
\tilde\alpha&{\sqrt{8\pi}\;\;\tilde \Gamma}_{\cal M}^{\tilde\alpha}\\[1ex]
\hline
^3S_1&{\hat \xi_{\cal M}}\\
^3D_1& -\frac{1}{\sqrt{2}}
\left[ {\hat \xi_{\cal M}}+\frac{3}{2}
({\hat k_1}-{\hat k_2})(p\xi_{\cal M})\nrmp^{-2}\right]\\
^3P_1& \sqrt{\frac{3}{2}}
\left[ \frac{1}{2} {\hat \xi_{\cal M}}({\hat k_1}-{\hat k_2})
-(p\xi_M) \right]\nrmp^{-1}\\
^1P_1&\sqrt{3} (p\xi_{\cal M})\nrmp^{-1}\\
\hline\hline
\end{array}
\]
\end{table}

\newpage
\widetext

\begin{center}
{\bf Figure captions}
\end{center}

\begin{figure}[ht]
\caption{
The nucleon density 
in the deuteron
computed within BS formalism in comparison with the
nonrelativistic results.}
\protect\label{chargedens}
\end{figure}

\begin{figure}[ht]
\caption{
The nucleon spin distribution
in the deuteron
computed within the BS formalism in comparison with the
nonrelativistic results.}
\protect\label{spindens}
\end{figure}

\begin{figure}[h]
\caption{
The momentum dependence of the $^3S_1^{++}$
component defined by eq.~(\protect\ref{wv1}) (solid line)
in  comparison with the corresponding nonrelativistic wave
functions with Bonn and Paris potentials 
(dotted and dashed lines, respectively.}
\label{sppkaz}
\end{figure}

\begin{figure}[h]
\caption{
The momentum dependence of the $^3D_1^{++}$
component. The
solid line (BS-I) depicts the result of computation by
(\protect\ref{wv1}); the dotted line (BS-II) includes
the contribution of $P$-waves
(see text); short- and long-dashed lines depict the
nonrelativistic wave
functions with Bonn and Paris potentials, respectively.}
\label{dppkaz}
\end{figure}

\begin{figure}[h]
\caption{
The momentum dependence of the $P$ waves defined by eq.~(\protect\ref{wv1})}.
\label{pkaz}
\end{figure}

\begin{figure}[h]
\caption{
The behavior
of the vertex function   $G(p_0,|{\bf p}|)$ for the $^3S_1^{++}$
configuration in the deuteron in dependence on $p_4$
and $|{\bf p}|.$}
\label{gspp3d}
\end{figure}

\begin{figure}[h]
\caption{
The same as fig.~\protect\ref{gspp3d}
but for the  $^3D_1^{++}$ configuration.}
\label{gdpp3d}
\end{figure}

\begin{figure}[h]
\caption{
The nonrelativistic limit of the $^3S_1^{++}$
component defined by eq.~(\protect\ref{sroed}) (solid line)
in comparison with the nonrelativistic wave
functions with Bonn and Paris potentials 
(dotted and dashed lines, respectively.}
\label{spgen}
\end{figure}

\begin{figure}[h]
\caption{
The same as fig. \protect\ref{spgen} but for
the $^3D_1^{++}$ components.}
\label{dpgen}
\end{figure}

\newpage
\phantom{.}
\vspace*{2cm}

\let\picnaturalsize=N
\def\picsize{12cm}
\def\picfilename{charge.ps}
\ifx\nopictures Y\else{\ifx\epsfloaded Y\else\input epsf \fi
\let\epsfloaded=Y
\centerline{\ifx\picnaturalsize N\epsfxsize
 \picsize\fi \epsfbox{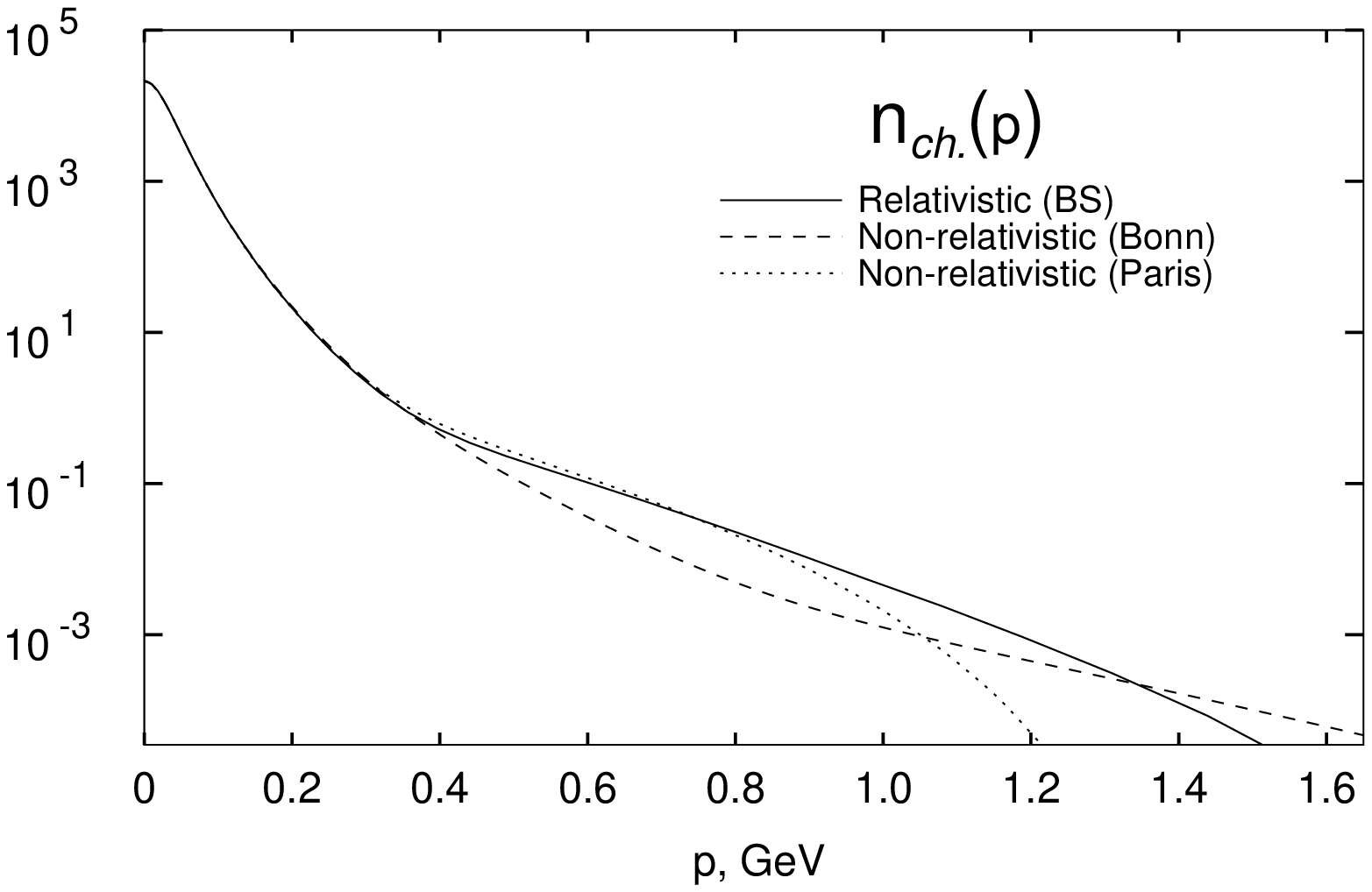}}}\fi

\vfill
Fig.~\ref{chargedens}. L.P. Kaptari, A. Umnikov.... Bethe-Salpeter Amplitudes...
\newpage
\phantom{.}
\vspace*{2cm}

\let\picnaturalsize=N
\def\picsize{12cm}
\def\picfilename{spin.ps}
\ifx\nopictures Y\else{\ifx\epsfloaded Y\else\input epsf \fi
\let\epsfloaded=Y
\centerline{\ifx\picnaturalsize N\epsfxsize
 \picsize\fi \epsfbox{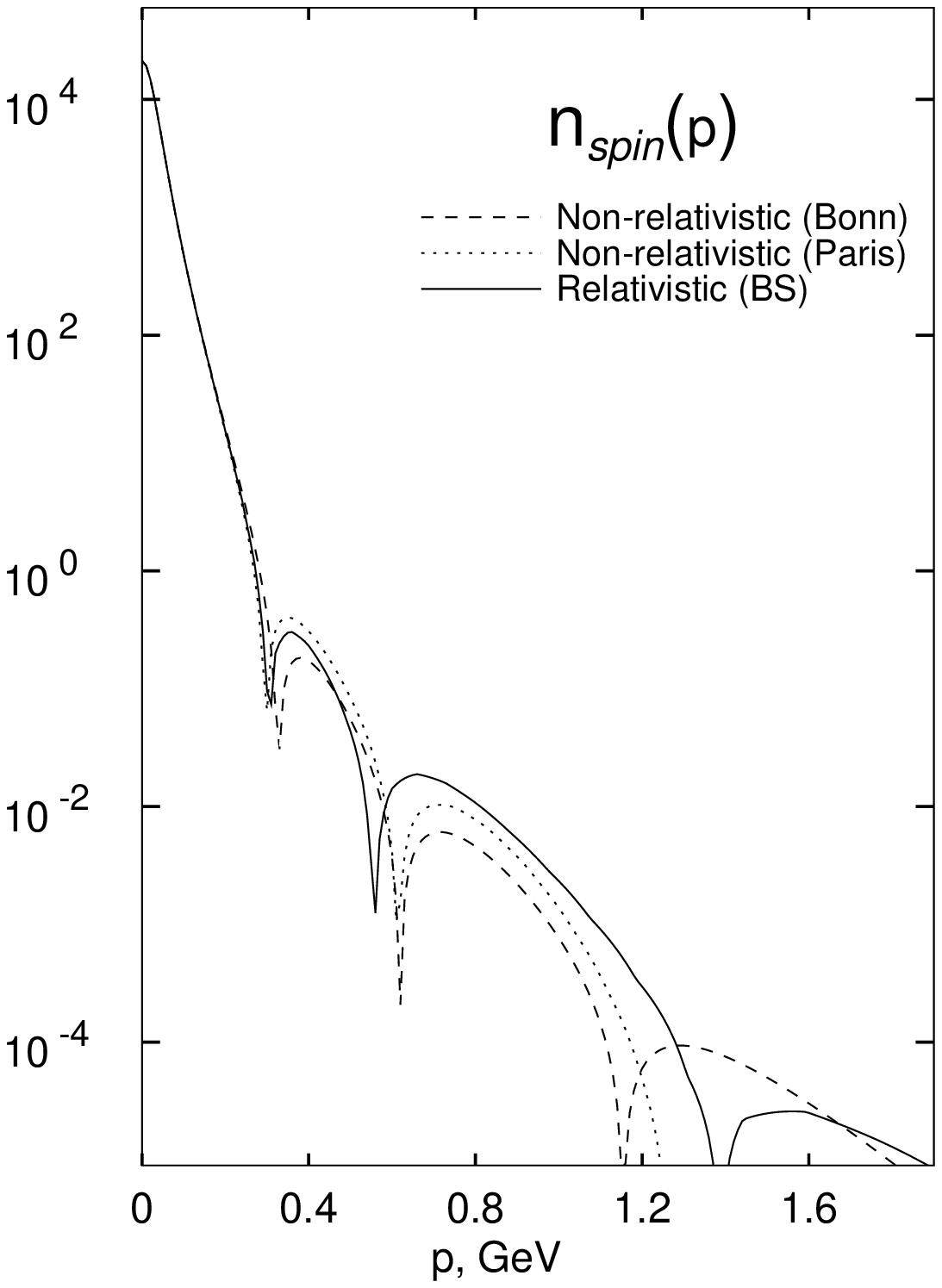}}}\fi

\vfill
Fig.~\ref{spindens}. L.P. Kaptari, A. Umnikov.... Bethe-Salpeter Amplitudes...
\newpage
\phantom{.}
\vspace*{2cm}

\let\picnaturalsize=N
\def\picsize{12cm}
\def\picfilename{sppkaz.ps}
\ifx\nopictures Y\else{\ifx\epsfloaded Y\else\input epsf \fi
\let\epsfloaded=Y
\centerline{\ifx\picnaturalsize N\epsfxsize
 \picsize\fi \epsfbox{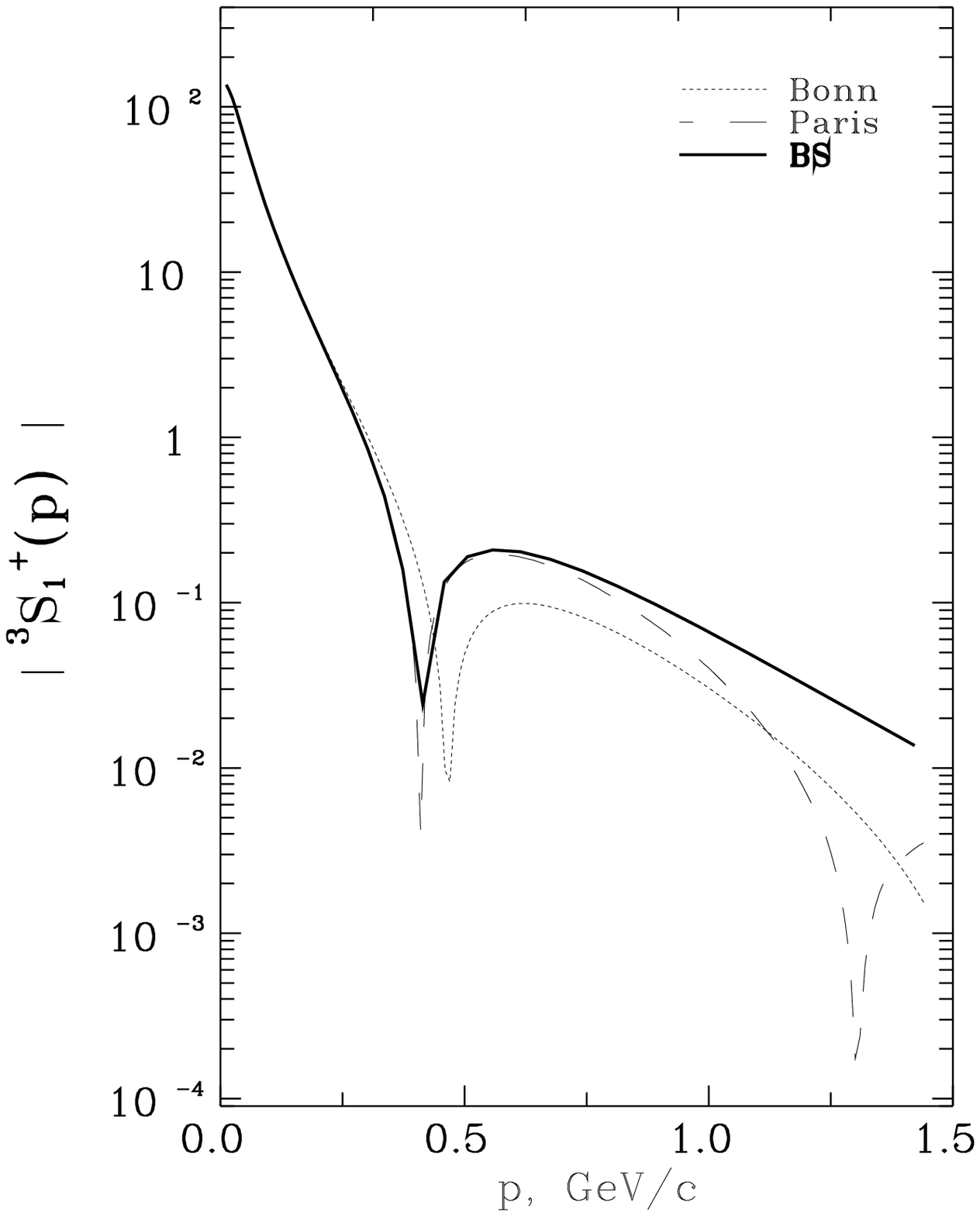}}}\fi

\vfill
Fig.~\ref{sppkaz}. L.P. Kaptari, A. Umnikov.... Bethe-Salpeter Amplitudes...

\newpage
\phantom{.}
\vspace*{2cm}

\let\picnaturalsize=N
\def\picsize{12cm}
\def\picfilename{dppkaz.ps}
\ifx\nopictures Y\else{\ifx\epsfloaded Y\else\input epsf \fi
\let\epsfloaded=Y
\centerline{\ifx\picnaturalsize N\epsfxsize
 \picsize\fi \epsfbox{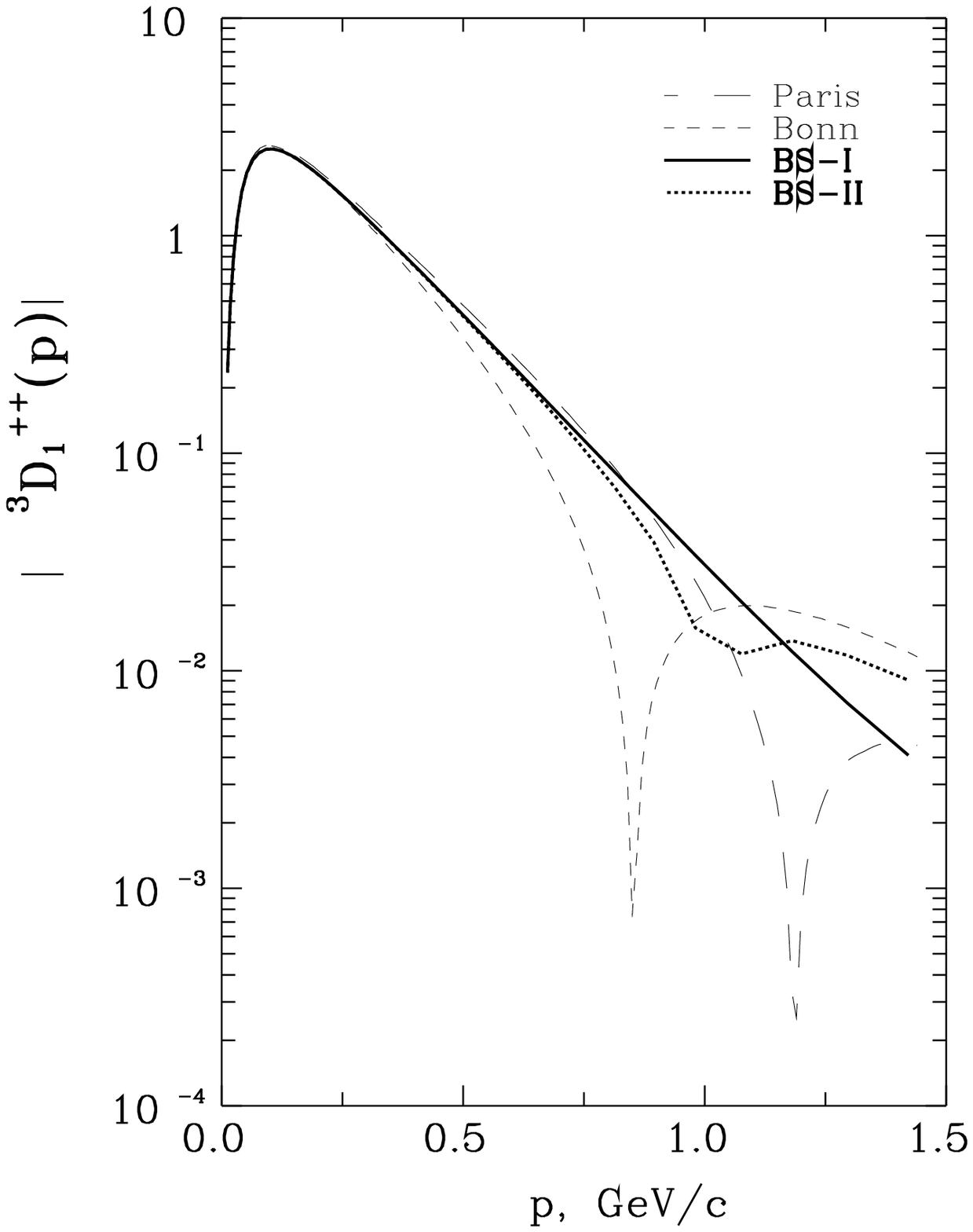}}}\fi

\vfill
Fig.~\ref{dppkaz}. L.P. Kaptari, A. Umnikov.... Bethe-Salpeter Amplitudes...

\newpage
\phantom{.}
\vspace*{2cm}

\let\picnaturalsize=N
\def\picsize{12cm}
\def\picfilename{pkaz.ps}
\ifx\nopictures Y\else{\ifx\epsfloaded Y\else\input epsf \fi
\let\epsfloaded=Y
\centerline{\ifx\picnaturalsize N\epsfxsize
 \picsize\fi \epsfbox{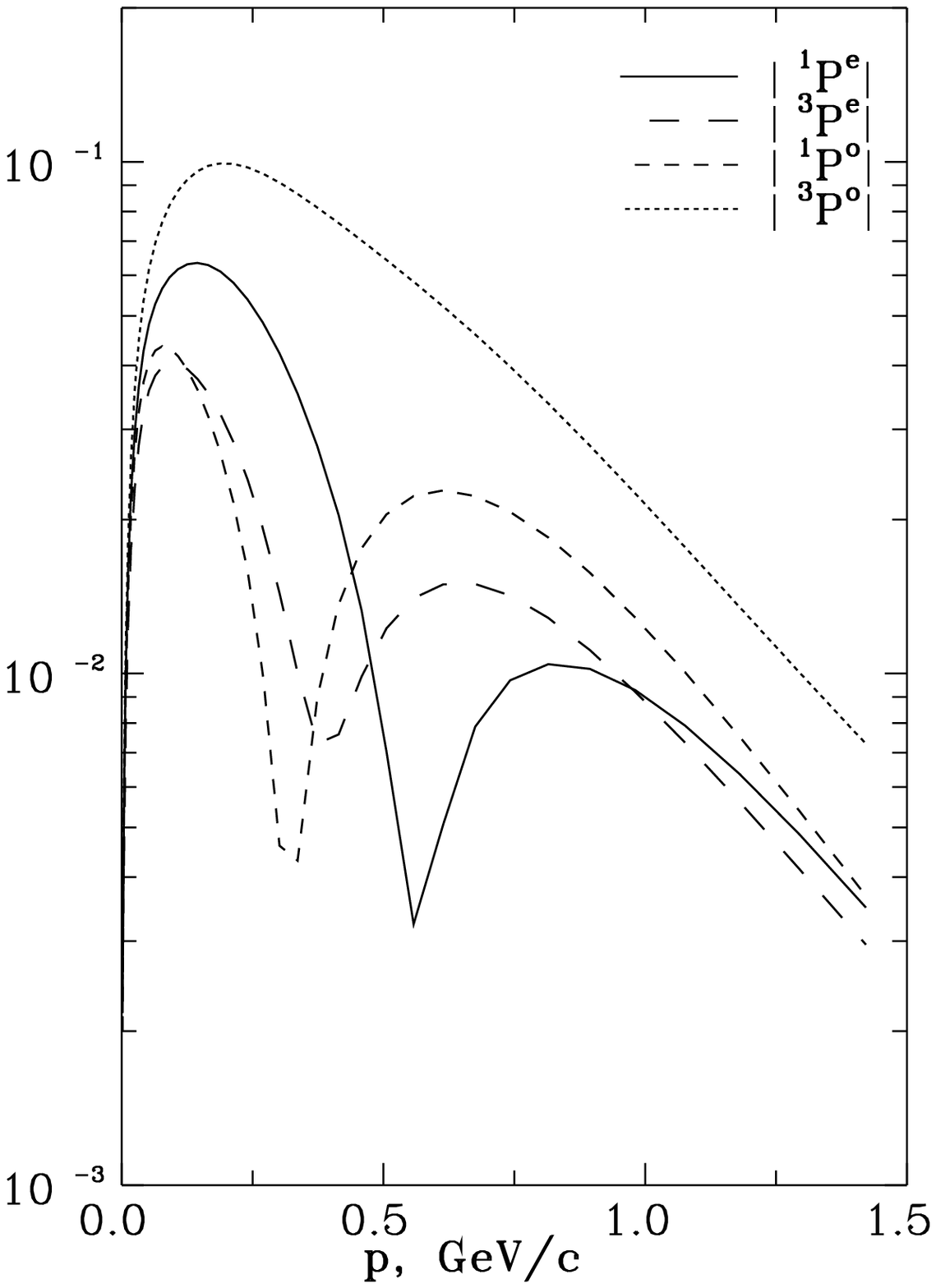}}}\fi

\vfill
Fig.~\ref{pkaz}. L.P. Kaptari, A. Umnikov.... Bethe-Salpeter Amplitudes...

\newpage
\phantom{.}
\vspace*{3cm}

\let\picnaturalsize=N
\def\picsize{14cm}
\def\picfilename{spp3d.ps}
\ifx\nopictures Y\else{\ifx\epsfloaded Y\else\input epsf \fi
\let\epsfloaded=Y
\centerline{\ifx\picnaturalsize N\epsfxsize
 \picsize\fi \epsfbox{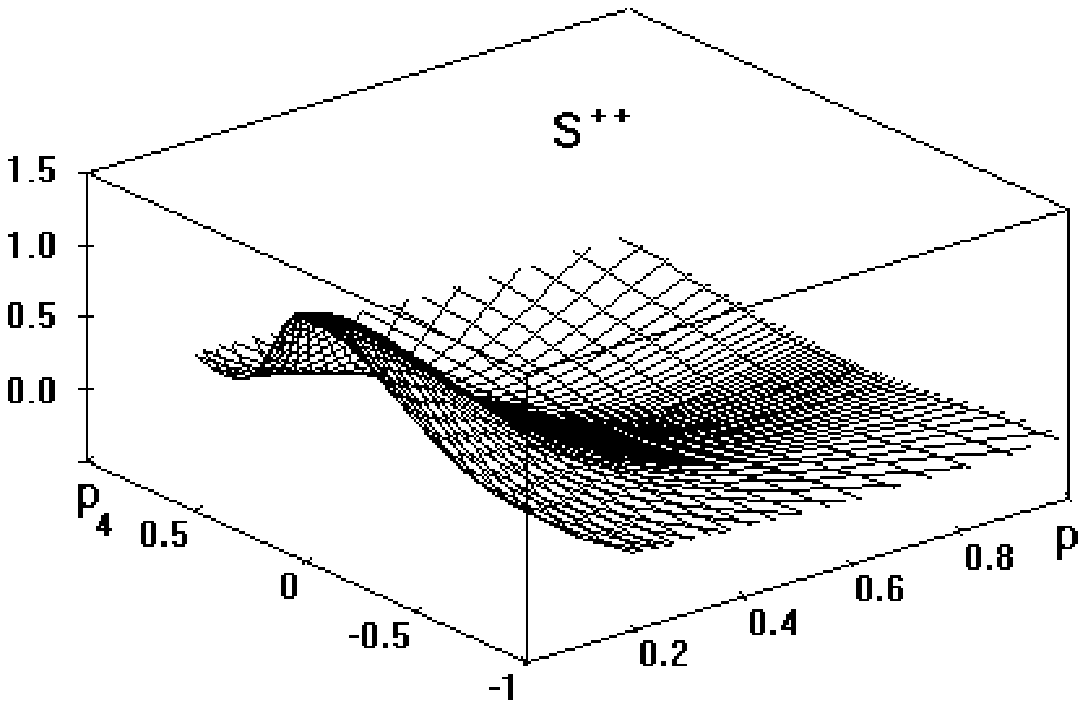}}}\fi

\vfill
Fig.~\ref{gspp3d}. L.P. Kaptari, A. Umnikov.... Bethe-Salpeter Amplitudes...
\newpage
\phantom{.}
\vspace*{3cm}

\let\picnaturalsize=N
\def\picsize{14cm}
\def\picfilename{dpp3d.ps}
\ifx\nopictures Y\else{\ifx\epsfloaded Y\else\input epsf \fi
\let\epsfloaded=Y
\centerline{\ifx\picnaturalsize N\epsfxsize
 \picsize\fi \epsfbox{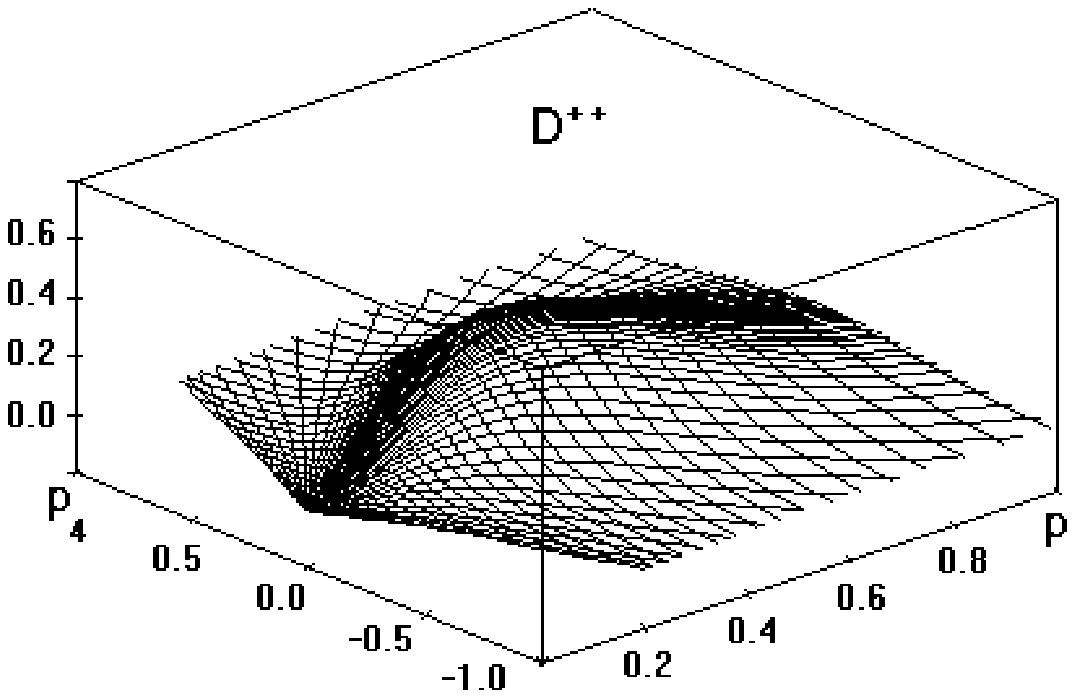}}}\fi

\vfill
Fig.~\ref{gdpp3d}. L.P. Kaptari, A. Umnikov.... Bethe-Salpeter Amplitudes...
\newpage
\phantom{.}
\vspace*{2cm}

\let\picnaturalsize=N
\def\picsize{12cm}
\def\picfilename{sppbs.ps}
\ifx\nopictures Y\else{\ifx\epsfloaded Y\else\input epsf \fi
\let\epsfloaded=Y
\centerline{\ifx\picnaturalsize N\epsfxsize
 \picsize\fi \epsfbox{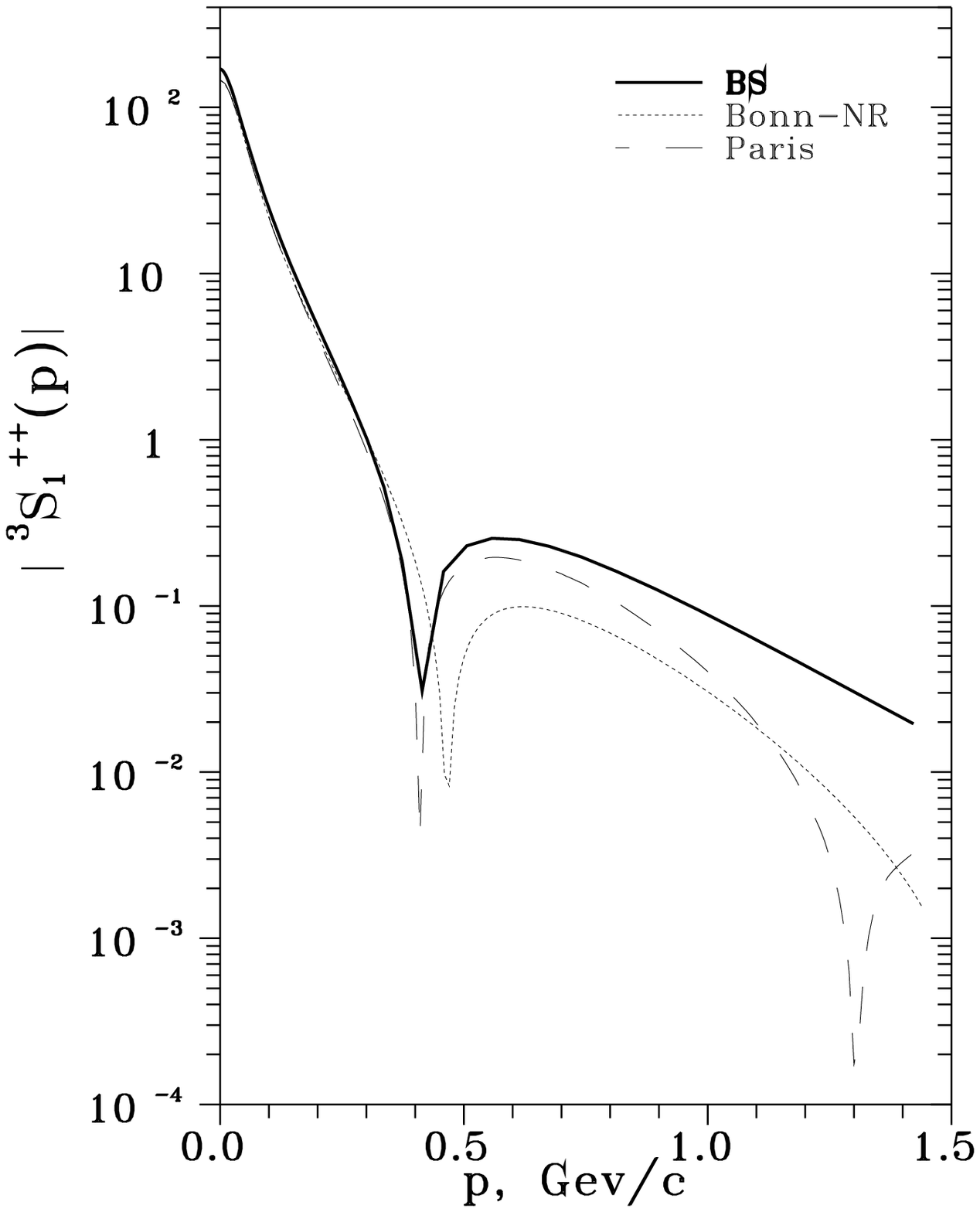}}}\fi

\vfill
Fig.~\ref{spgen}. L.P. Kaptari, A. Umnikov.... Bethe-Salpeter Amplitudes...
\newpage
\phantom{.}
\vspace*{2cm}

\let\picnaturalsize=N
\def\picsize{12cm}
\def\picfilename{dppbs.ps}
\ifx\nopictures Y\else{\ifx\epsfloaded Y\else\input epsf \fi
\let\epsfloaded=Y
\centerline{\ifx\picnaturalsize N\epsfxsize
 \picsize\fi \epsfbox{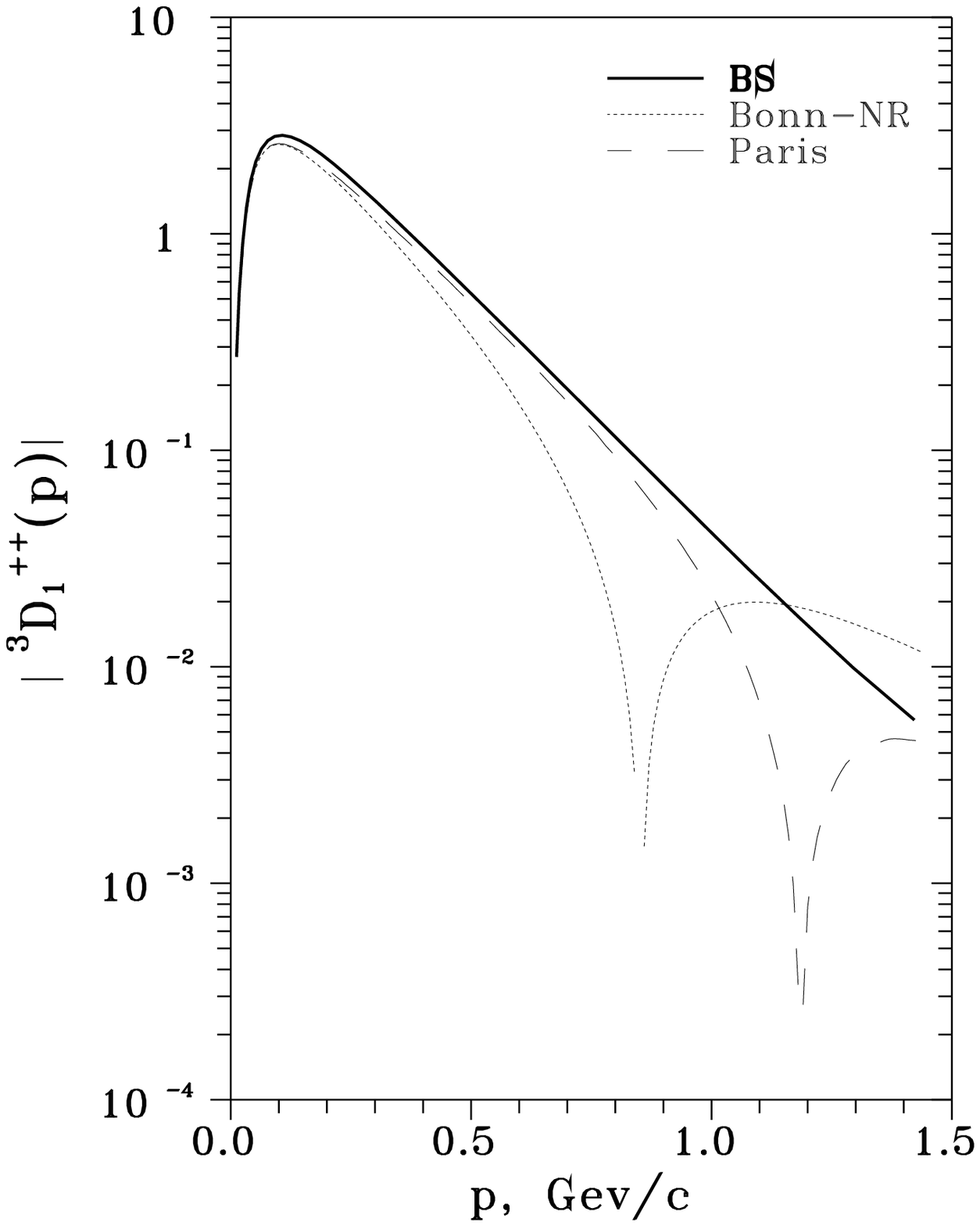}}}\fi

\vfill
Fig.~\ref{dpgen}. L.P. Kaptari, A. Umnikov.... Bethe-Salpeter Amplitudes...
\newpage

\end{document}